\documentclass[fleqn,usenatbib]{mnras}

\usepackage{newtxtext,newtxmath}

\usepackage[T1]{fontenc}

\DeclareRobustCommand{\VAN}[3]{#2}
\let\VANthebibliography\thebibliography
\def\thebibliography{\DeclareRobustCommand{\VAN}[3]{##3}\VANthebibliography}

\usepackage{graphicx}	
\usepackage{amsmath}	
\newcommand{\sersic}{S\'ersic}

\usepackage{indentfirst}
\usepackage{appendix}
\usepackage{makecell}
\usepackage{tabu}
\usepackage{float}
\usepackage{xcolor}
\usepackage{amsmath}
\usepackage{pdflscape}
\usepackage{cleveref}
\DeclareUnicodeCharacter{2212}{-}
\DeclareUnicodeCharacter{00B4}{'}
\DeclareUnicodeCharacter{00A8}{"}

\title[Quasar host asymmetries]{Morphological asymmetries of quasar host galaxies with Subaru Hyper Suprime-Cam}

\author[S. Tang et al.]{
Shenli Tang,$^{1,2,3,4}$\thanks{E-mail: tang-shenli897@g.ecc.u-tokyo.ac.jp}
John D. Silverman,$^{2,3,5}$
Hassen M. Yesuf,$^{2,6}$
Xuheng Ding,$^{2}$
Junyao Li,$^{7}$
Connor Bottrell,$^{2}$\newauthor
\textcolor{black}{Andy Goulding,$^{8}$}
Kiyoaki Christopher Omori,$^{9}$
Yoshiki Toba,$^{10,11,12}$
and Toshihiro Kawaguchi$^{13}$
\\
$^{1}$Department of Physics, School of Science, The University of Tokyo, 7-3-1 Hongo, Bunkyo-ku, Tokyo 113-0033, Japan\\
$^{2}$Kavli Institute for the Physics and Mathematics of the Universe (WPI), The University of Tokyo, Kashiwa, Chiba 277-8583, Japan\\
$^{3}$Center for Data-Driven Discovery, Kavli IPMU (WPI), UTIAS, The University of Tokyo, Kashiwa, Chiba 277-8583, Japan\\
$^{4}$Institute for Cosmic Ray Research, The University of Tokyo, 5-1-5 Kashiwanoha, Kashiwa, Chiba 277-8582, Japan\\
$^{5}$Department of Astronomy, School of Science, The University of Tokyo, 7-3-1 Hongo, Bunkyo, Tokyo 113-0033, Japan\\
$^{6}$Kavli Institute for Astronomy and Astrophysics, Peking University, Beijing 100871, China\\
$^{7}$Department of Astronomy, University of Illinois at Urbana-Champaign, Urbana, IL 61801, USA\\
$^{8}$Department of Astrophysical Sciences, Princeton University, Princeton, NJ 08540, USA\\
$^{9}$Division of Particle and Astrophysical Science, Nagoya University, Furo-cho, Chikusa-ku, Nagoya 464–8602, Japan\\
$^{10}$National Astronomical Observatory of Japan, 2-21-1 Osawa, Mitaka, Tokyo 181-8588, Japan\\
$^{11}$Academia Sinica Institute of Astronomy and Astrophysics, 11F of Astronomy-Mathematics Building, AS/NTU, No.1, Section 4, Roosevelt Road, Taipei 10617, Taiwan\\
$^{12}$Research Center for Space and Cosmic Evolution, Ehime University, 2-5 Bunkyo-cho, Matsuyama, Ehime 790-8577, Japan\\
$^{13}$Department of Economics, Management and Information Science, Onomichi City University, Hisayamada 1600-2, Onomichi, Hiroshima 722- 8506, Japan
}

\date{Accepted XXX. Received YYY; in original form ZZZ}

\pubyear{2023}

\begin{document}
\label{firstpage}
\pagerange{\pageref{firstpage}--\pageref{lastpage}}
\maketitle

\begin{abstract}
How does the host galaxy morphology influence a central quasar or vice versa? We address this question by measuring the asymmetries of 2424 SDSS quasar hosts at $0.2<z<0.8$ using broad-band ($grizy$) images from the Hyper Suprime-Cam Subaru Strategic Program. Control galaxies (without quasars) are selected by matching the redshifts and stellar masses of the quasar hosts. A two-step pipeline is run to decompose the PSF and \sersic\ components, and then measure asymmetry indices ($A_{\rm CAS}$, $A_{\rm outer}$, and $A_{\rm shape}$) of each quasar host and control galaxy. \textcolor{black}{We find a mild correlation between host asymmetry and AGN bolometric luminosity ($L_{\rm bol}$) for the full sample (spearman correlation of 0.37) while a stronger trend is evident at the highest luminosities ($L_{\rm bol}>45$). This then manifests itself into quasar hosts being more asymmetric, on average, when they harbor a more massive and highly accreting black hole. The merger fraction also positively correlates with $L_{\rm bol}$ and reaches up to 35\% for the most luminous. Compared to control galaxies, quasar hosts are marginally more asymmetric (excess of 0.017 in median at 9.4$\sigma$ level) and the merger fractions are similar ($\sim 16.5\%$). We quantify the dependence of asymmetry on optical band which demonstrates that mergers are more likely to be identified with the bluer bands and the correlation between $L_{\rm bol}$ and asymmetry is also stronger in such bands. We stress that the band dependence, indicative of a changing stellar population, is an important factor in considering the influence of mergers on AGN activity.} 
\end{abstract}




\section{Introduction}
Measurements of morphology and structural parameters of type-1 quasar hosts are hampered due to the central bright point source. However, these measurements are essential to understand the role that supermassive black holes (SMBHs) play in galaxy evolution. It is suggested that the merger of galaxies may drive gas inflow onto the central SMBH, thus igniting a luminous quasar (active galactic nuclei (AGNs) with $L_{\text {bol}} \gtrsim 10^{43}$ erg $\mathrm{s}^{-1}$) \citep{hernquist1989tidal,barnes1991fueling,di2005energy,capelo2015growth}, and at the same time, build up the tight correlations between the BH mass and stellar velocity dispersion \citep{ferrarese2000fundamental, gebhardt2000relationship, merritt2001black} and bulge mass \citep{laor2001linearity, mclure2002black, marconi2003relation, haring2004black}. In this scenario, it is expected that AGNs show more disturbed features than inactive galaxies. \par
A traditional way to test this scenario is to select a control inactive galaxy sample and visually classify the mergers in both populations \citep{cisternas2010bulk,kocevski2011candels,weston2016incidence,fan2016most,mechtley2016most,ricci2017growing,donley2018evidence,ellison2019definitive,marian2019major,marian2020significant}. However, visual classification is subjective and difficult to reproduce. Additionally, quantitative and systematic approaches to merger classification are necessary when handling large data volumes from current and future astronomical imaging surveys. \par
\cite{conselice2003relationship} developed an asymmetry index (A) to classify major mergers and showed that ULIRGS tend to have asymmetry indices larger than 0.35, which was further used to identify major mergers out to $z\sim3$ in HDF. The A index is measured by rotating the image of a galaxy by $180^{\circ}$ and subtract it from the original image. After being normalized by the total flux of the galaxy, this value can well describe the asymmetry of a galaxy \citep{elmegreen1982flocculent,conselice1997symmetry}. Besides mergers, asymmetry can also be used as an indicator of star-formation activity. \citep[e.g.,][]{conselice2003relationship,conselice2014evolution,yesuf2021important}. Since blue star-forming galaxies tend to be more asymmetric than red quiescent galaxies. \par
Regarding quasars, only a few studies measured A index of of their host galaxies due to the difficulty of separating the contribution from the quasar light from its host and minimizing its effect on the asymmetry measurements \citep{zhao2022relation}. Consequently, most studies have focused on Type-2 AGNs \citep{gabor2009active,boehm2013agn,lackner2014late,glikman2015major,goulding2018galaxy,zhao2022relation}, and some mixed the two populations with X-ray selection \citep{kocevski2011candels,cotini2013merger,villforth2014morphologies,villarroel2017agn}. 

In particular, \cite{goulding2018galaxy} studied the morphology of 2552 WISE-selected AGNs at $z<0.9$ in Subaru HSC-SSP footprint. The merger fraction was then estimated using a Random Forest Classifier based on the morphological parameters including asymmetry. They found that the AGN fraction in merger samples are a factor of $2\sim7$ higher than in non-interacting galaxies, and the merger fraction of AGNs is enhanced in the most luminous population. Their results strongly suggest a positive connection between the mergers and AGNs. However, they excluded Type-1 AGNs from their samples due to the stronger contamination from the point sources. Studies have pointed out that Type-1 AGNs have intrinsic differences with Type-2 AGNs in their physical properties \citep{satyapal2014galaxy,zakamska2016star,villarroel2017agn,kong2018black,zhuang2020interplay}, such as Eddington ratio and star formation rate (SFR). Also, how heavily the AGNs are obscured may also make a factor of 2 difference in merger fraction \citep{kocevski2015compton}, but currently tested with only $\sim150$ sources. Therefore, a dedicated effort focusing on Type-1 AGNs with comparable statistics to the  \cite{goulding2018galaxy} study is required to fill in the gap. Furthermore, Type-1 AGNs will allow us to investigate the dependence of merger features on BH mass and Eddington ratio.
\par
While there is growing consensus that mergers are not the main driver of BH accretion, there remains some discrepancies on how much of an excess in the merger fraction of quasars exists compared to inactive galaxies. X-ray studies typically find no excess of the AGN merger fraction over inactive galaxies at $0.3\lesssim z\lesssim 2.5$ \citep{gabor2009active,kocevski2011candels,boehm2013agn,villforth2014morphologies,villarroel2017agn}; however, \citet{silverman2011impact} 
 find a $2\times$ enhancement at $0.2 < z < 1.0 $ and 
\cite{cotini2013merger} report a $\sim 5\times$ excess in the local Universe. Optical selected Type-1 AGNs show observational evidence for both positive \citep{ellison2011galaxy,marian2020significant} and negative \citep{cisternas2010bulk,marian2019major,zhao2022relation} excess for the merger fraction. The variation in AGN merger fractions, as exhibited in the aforementioned studies, is likely attributed to a number of factors including luminosity evolution \citep{treister2012major,weigel2018fraction,kim2021hubble}, redshift evolution \citep{steinborn2018cosmological}, and merger-stage dependence \citep{ricci2017growing,goulding2018galaxy}. 

In this work, we present another approach to lessen the tension between various measurements by considering the properties of the stellar population of AGN host galaxies. We measure the asymmetry indices from five broad-band images, which allows us to compare asymmetry as a function of wavelength which differs due to the contributions from different stellar populations. In addition, we study correlations between physical properties of our quasars and asymmetries. Such correlations could be triggered by the dependence of merger fractions on those properties as mentioned above.\par

Here, we use 2424 Sloan Digital Sky Survey \citep[SDSS][]{york2000aj} Type-1 quasars at $0.2<z<0.8$ selected from the DR14 quasar catalog \citep{myers2015sdss,paris2018sloan}, with measured properties (sizes, \sersic\ indices) of their quasar hosts from \cite{li2021sizes} based on Subaru imaging from the HSC-SSP \citep[][s21a\_wide]{miyazaki2018hyper}. High-quality optical imaging enables parametric morphological decompositions to be more accurate, and improve measurements of galaxy asymmetries, as shown by \cite{bottrell2019bulge} using deep and shallow images from SDSS Stripe 82. The wide coverage of HSC-SSP allows us to build the largest statistical sample of Type-1 quasar hosts for non-parametric asymmetry measurements. A common practice in previous works is to measure quasar host asymmetry using the \textsc{Galfit} \citep{peng2002detailed} package to decompose the quasar image and remove the point source. Then the asymmetry is calculated on the $pure$ quasar host image \citep[e.g.,]{zhao2022relation}. In this work, we follow this routine, while making use of a more recently developed decomposition tool \textsc{GaLight} \citep{ding2021galaxy}, and morphology measurement tool \textsc{statmorph} \citep{rodriguez2019optical}.\par

This paper is arranged as following: \cref{sec:selection} describes the sample selection. \cref{sec:methods} presents the methodology of measurements. Results are shown in \cref{sec:results}, and further discussion are made in \cref{sec:discussion}. We adopt $\Lambda CDM$ cosmology with $\Omega_{\Lambda}=0.7$, $\Omega_{\mathrm{m}}=0.3$, and $H_{0}=70 \mathrm{~km} \mathrm{~s}^{-1} \mathrm{Mpc}^{-1}$ across the paper.


\begin{figure*}
\begin{centering}
\includegraphics[width=0.9\textwidth]{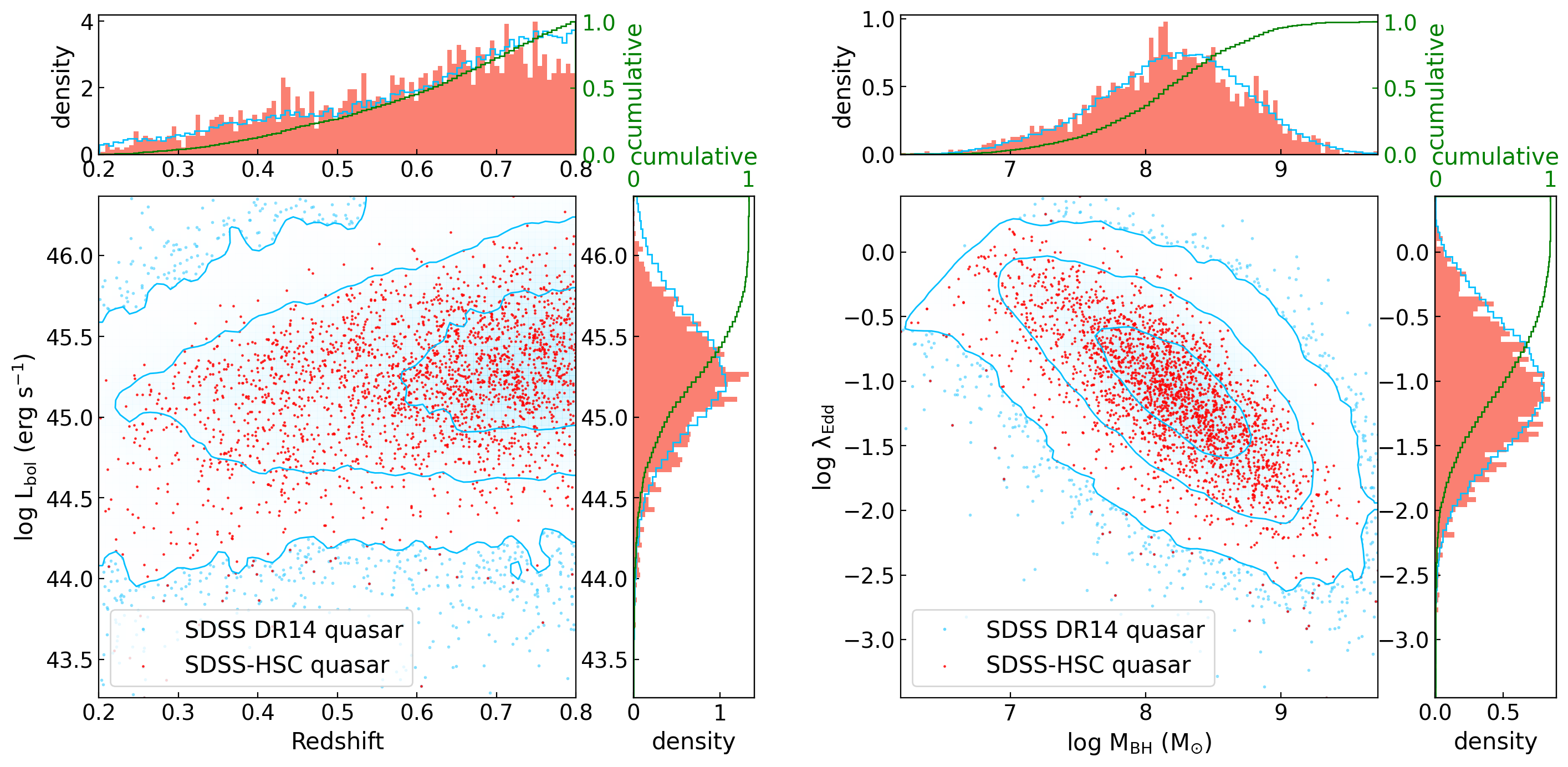}
\end{centering}
\caption{\textcolor{black}{Left: $L_{\rm bol}$ as a function of redshift for Type-1 quasars used in this work (red points in the main plot and red histograms in the appended plots) and the entire SDSS DR14 quasar population in the same redshift range (blue contours in the main plot and blue empty histograms in the appended plots). The green cumulative histograms show the HSC quasar distribution. Right: Similar to the left, but for Eddington ratio and $M_{\rm BH}$. \label{fig:BH_edd}}}
\end{figure*}

\section{Sample selection} \label{sec:selection}

The optical imaging used in this study is from the HSC-SSP data release 4 (internal). It covers $\sim 1100$ deg$^2$ with five broad bands \citep[$grizy$][]{kawanomoto2018hyper} to a full $5\sigma$ depth reaching 26.5, 26.5, 26.2, 25.2, 24.4 mag with median seeing of 0.79, 0.75, 0.61, 0.68, 0.68 arcseconds respectively \citep{aihara2022third}. The data were reduced with the HSC Pipeline \citep{bosch2018hyper}.

\subsection{Type-1 quasars} \label{subsec:Q_selection}

\textcolor{black}{The type-1 quasars used in the \cite{li2021sizes} study} were originally selected from SDSS DR14 quasar catalog \citep{myers2015sdss,paris2018sloan} \textcolor{black}{and cross-matched} with HSC-SSP PDR2 \citep{aihara2019second}. The authors studied the host galaxy size-mass relation and $M_{\rm BH}$ - $M_*$ mass relation of these quasars \citep{li2021synchronized}. The final sample contains 2424 quasars at $0.2<z<0.8$. The lower limit was set to avoid fine structures such as spiral arms that could not be well modeled in the parametric fitting. We keep this lower limit because we are interested in the correlations between those parameters and asymmetry. While at $z>0.8$, the host galaxies are difficult to detect in HSC due to the surface brightness dimming \citep{ishino2020subaru}. \textcolor{black}{The quasar host galaxies are magnitude-limited ($i_{mag}<23$) and have 90\% completeness at lower limits of $10^{9.3-10.3} M_{\odot}$ for z=0.3-0.7. This selection results in 13.5\% of the sources being rejected by these limits. An upper limit of $10^{11.5} M_{\odot}$ was set because galaxies above this threshold are very likely undetected in blue bands thus may have inaccurate stellar mass measurements. Such a cut may lead to a biased fraction towards star-forming hosts among the quasars, but this number is minimal ($\sim1\%$). They also applied some quality checks to the subtraction and fitting results as detailed in Section 5 and Table 2 of their paper.} In this work, we updated the imaging from PDR2 to DR4 s21a\_wide, because the sky subtraction algorithm was optimized over the recent data release. In particular, the global sky extraction was improved to deal with a bias that might be caused by dead CCDs \citep[see Figure 7 of ][]{aihara2022third}. \textcolor{black}{In addition, local sky subtraction was implemented to mitigate the impact of extended wings of bright objects. We note that a complementary paper (Yesuf et al. in prep) will examine the asymmetries of a large sample of BLAGNs at $z<0.35$ imaged by HSC.} 
\par
The BH properties and redshifts of these quasars are estimated by \cite{rakshit2020spectral} based on SDSS spectral properties. Specifically, BH masses ($M_{\rm BH}$) are measured from the single epoch virial method using H$\beta$ and $L_{5100}$ \citep{vestergaard2006determining}:
\begin{equation}
M_{\mathrm{BH}}=10^{6.91} L^{0.5}_{5100,44}\left(\frac{\mathrm{FWHM}_{H\beta}}{1000 \mathrm{~km} \mathrm{~s}^{-1}}\right)^{2} M_{\odot}
\end{equation}
where $L_{5100,44}$ is the \textcolor{black}{monochromatic} luminosity at rest-frame 5100 \text{\AA} in units of $10^{44}\ \rm{erg\ s^{-1}}$ \textcolor{black}{measured by power-law continuum fitting to the SDSS spectra}. The bolometric luminosity $L_{\rm bol}$ is calculated using the bolometric correction factor in \cite{richards2006spectral}:
\begin{equation} \label{eq:Lbol}
L_{\mathrm{bol}} = 9.26 \times L_{5100}
\end{equation}
Then the Eddington ratio is given by:
\begin{equation} \label{eq:R_edd}
\lambda_{\rm edd} = \log (L_{\mathrm{bol}}/L_{\mathrm{edd}})
\end{equation}
where $L_{\mathrm{edd}}$ is the Eddington luminosity estimated from $M_{\rm BH}$:
\begin{equation} \label{eq:L_edd}
L_{\mathrm{Edd}} \cong 1.3 \times 10^{38}\left(M_{\mathrm{BH}} / M_{\odot}\right) \mathrm{erg} \mathrm{s}^{-1}
\end{equation}

\textcolor{black}{We show the distribution of the 2424 quasars used in this work in $L_{\rm bol}$ as versus redshift and $\lambda_{\rm edd}$ versus $M_{\rm BH}$ diagrams (Figure \ref{fig:BH_edd}).  Compared to the whole SDSS DR14 quasar population, our samples are slighted shifted to low $L_{\rm bol}$, thus the low $M_{\rm BH}$ and low $\lambda_{\rm edd}$ region. There is also a small decrement in density ($\sim30\%$) at $z>0.75$. Both of the above biases are caused by the failure in detecting the host galaxies of these populations, i.e., relatively bright and high-z quasars being rejected in \cite{li2021sizes} catalog.}\par

\subsection{Control Galaxies} \label{subsec:control}
Control galaxies are selected from the catalog of \cite{kawinwanichakij2021hyper}, which includes $\sim$ 1.8 million galaxies in HSC PDR2 footprints with single \sersic\ fits. We also updated the images to DR4 data for the same reason as mentioned above. \cite{kawinwanichakij2021hyper} controlled the quality of galaxy images by setting the following parameters in the HSC-SSP database:
\begin{enumerate}
\item (g\&r)\_inputcount\_value $\geq$ 4
\item (i\&z\&y)\_inputcount\_value $\geq$ 6
\item i\_cmodel\_mag brighter than 24.5 AB
\item i\_cmodel\_magerr $\leq$ 0.1 AB
\item i\_extendedness\_value = 1 
\item PSFMAG−CMODELMAG $>$ 0.2
\end{enumerate}

\textcolor{black}{In addition}, we increased the threshold of the 6th criterion from the original value of 0.145 to 0.2 to be more strict with the removal of point sources. We find that some close double stars could be included when using the lower value. \textcolor{black}{To assure that our control sample does not include AGNs, we matched our control galaxies with the latest Million Quasars (Milliquas) catalog v7.8c \citep{flesch2021million}, which includes 1.4 million type-1, type-2 quasars selected from radio, optical, and X-ray surveys, and the HSC quasar catalog, which is based on HSC broad band color selections (Goulding et al. in prep.). This rejects 0.02\% and 0.04\% sources from our control galaxies, respectively.} Stellar mass and photo-$z$ of these galaxies, published in \cite{nishizawa2020photometric}, are estimated using a template fitting-code \textsc{mizuki} \citep{tanaka2015photometric, tanaka2018photometric}. The code uses a set of theoretical templates generated with the \cite{bruzual2003stellar} stellar population synthesis code to fit the CModel magnitudes of the sources. The results are calibrated to spectroscopic redshifts achieved from other surveys that overlap with the HSC-SSP survey. Stellar mass is estimated as the median value derived from the PDF marginalized over all the other parameters including SFR and dust attenuation. \par
\begin{figure}
\begin{centering}
\includegraphics[width=0.45\textwidth]{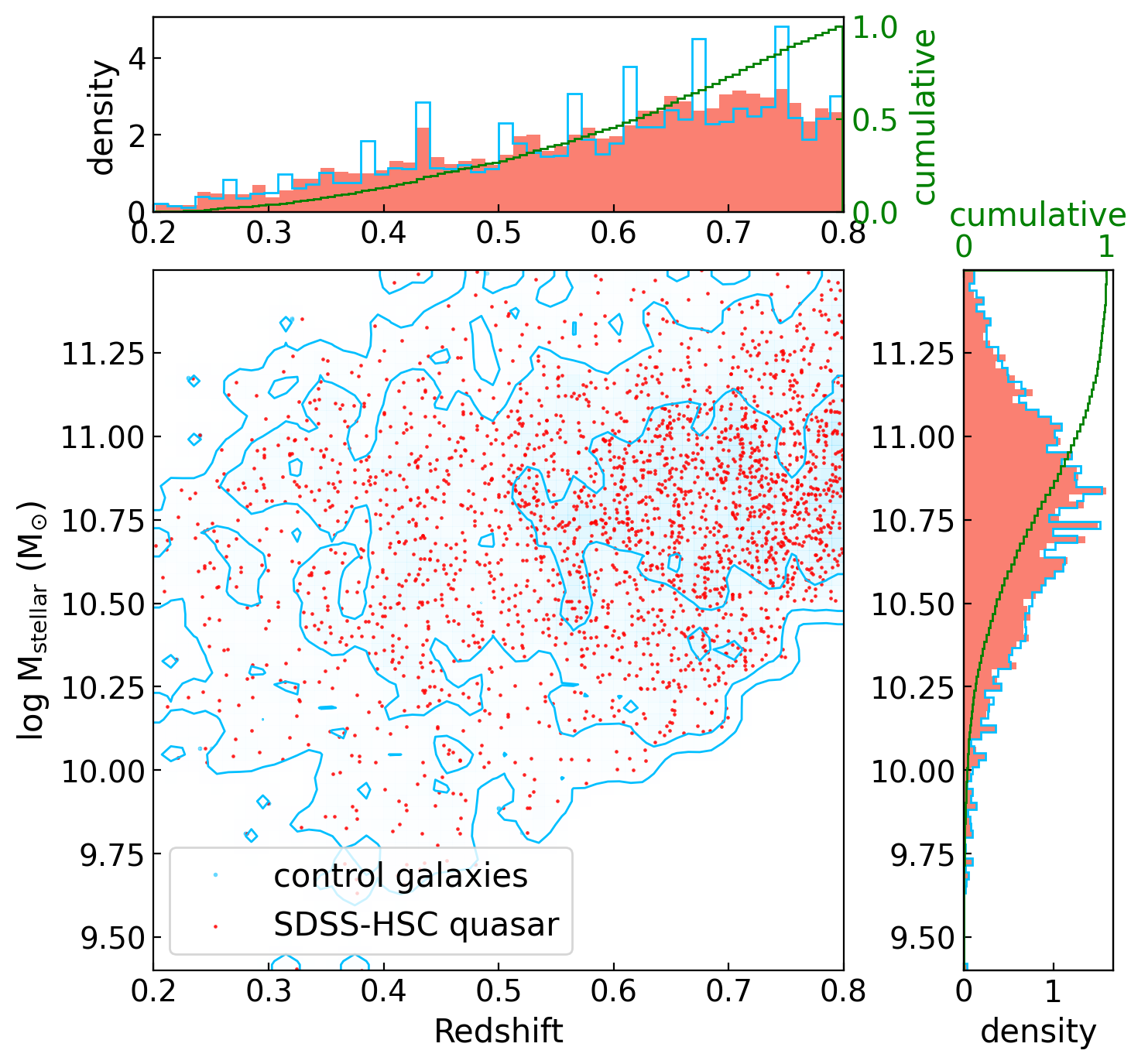}
\end{centering}
\caption{\textcolor{black}{The distribution of stellar mass and redshift for quasar hosts (red dots and red histograms) and control galaxies (blue contours and blue empty histograms). The peaked redshifts of control galaxies at certain values are artifacts of the binning and precision of photo-$z$. Green curves are the cumulative distribution for the quasars hosts.} \label{fig:QG_compare}}
\end{figure}

Based on these measurements, we apply the KD-Tree method to select control galaxies matching our quasars in redshift and stellar mass. KD-Tree is a binary space partitioning method that can efficiently find neighborhood data points in k-dimensional space \citep{fuchs1980visible}. We set the leaf size to 20 in \textsc{sklearn.neighbors.KDTree} and select 3 inactive galaxies for every quasar. We allow duplicated selection of galaxies, such galaxies are counted multiple times in our results. In total, \textcolor{black}{7060} galaxies are selected, with \textcolor{black}{212} duplicates. We show the redshift and stellar mass distribution of these quasars and control galaxies in Figure \ref{fig:QG_compare}. \par
\textcolor{black}{The photo-$z$ measurements of the control galaxies were qualified by \cite{kawinwanichakij2021hyper} as shown in their Appendix A, Figure 12. They compared HSC-SSP PDR2 \textsc{mizuki} photo-$z$ with available spec-$z$ measurements from other surveys such as zCOSMOS \citep{lilly2009zcosmos}, VIPERS \citep{garilli2014vimos} and PRIMUS \citep{cool2013prism}. They reported that the median offset between the spec-$z$ and photo-$z$ is 0.001. While $\sim 7\%$ of the sample has catastrophic failures ($|\Delta z| /\left(1+z_{\text {spec }}\right)>0.15$), most of which happen at $z>1.0$ or $z<0.2$. Between this redshift range, the photo-$z$ is well-constrained thanks to the 4000 \text{\AA} break, in which the failure fraction reduces to 3\%. We expect this fraction to be lower for our control galaxies as they are relatively bright sources (typical $i$-band magnitudes $\sim20.7$ compared to the lower cut of 24.5 in \cite{kawinwanichakij2021hyper}). On the other hand, the authors expect a typical scatter $\sim 0.25$ dex for the \textsc{mizuki} stellar mass measurements. They applied correction of $0.08-0.11$ dex for the ``out-shining” effect of star-forming galaxies \citep{sorba2015missing}. The uncertainties on the stellar mass of quasar hosts were estimated by \cite{li2021sizes} via comparison to simulated quasar images based on HST images in CANDELS-COSMOS field. The values weakly depend on masses and are typically between 0.1-0.3 dex.} \par

\section{Methodology} \label{sec:methods}

We measure the asymmetry of the quasar host galaxies and a matched control sample. For the former, this requires us to generate an image of the hosts free of quasar emission. Thus, we decompose the optical images into separate source components (i.e., quasar and host emission; Section \ref{subsec:decomposition}). As detailed below, two images are constructed, one with the nearby companions and the other without. This allows us to assess the level of asymmetric features either external or internal to quasar hosts (Section \ref{subsec:morph_measure}). The procedures for control galaxies are similar with details provided in Appendix \ref{sec:control_G}.

\subsection{Quasar subtraction through 2D decomposition} \label{subsec:decomposition}

We use \textsc{GaLight} \citep{ding2020mass,ding2021galaxy} to perform two-dimensional decomposition of quasar and host emission. \textsc{GaLight} is developed under Python3 environment and incorporates structures from \textsc{Lenstronomy} \citep{birrer2021lenstronomy}. For example, the particle swarm optimization \citep[PSO; ][]{kennedy1995particle} algorithm used in \textsc{GaLight} offers improvements in the gradient descent algorithms used in \textcolor{black}{other analysis tools which can become trapped in local minimal and dependent on the initial input values. \textsc{GaLight} models the imaging in multiple steps. First, the cutout size of the image frame is determined that includes any extended emission (e.g., tidal tails) of nearby objects, if present, and minimizes computing time by limiting the total number of pixels.} Square cutouts, centered on the target, have dimensions ranging from 84 to 196 pixels ($14-33\arcsec$ in HSC) in steps of 28 pixels. \par

We then fit all galaxies detected in the cutout image using \sersic\ profiles. We set $supersampling\_factor$ to 3 instead of 1 to allow a higher resolution sub-pixel sampling of surface brightness. The quasar component is represented by an additional point spread function (PSF) model which is generated by HSC Pipeline at the same position as the quasar \citep{bosch2018hyper}. Lastly, \textsc{GaLight} uses the PSO algorithm with 200 iterations to minimize the overall $\chi^2$ for the model parameters including \sersic\ radius, \sersic\ index, ellipticity, and central positions of the sources. In this work, we fixed the  \sersic\ radius and \sersic\ index to the same as measured in \cite{li2021sizes} to better reproduce the host magnitudes. Thus, their SED fitting results (e.g., stellar mass and rest frame U-V color) can be directly used in this work.\par

\begin{figure*}
\begin{centering}
\includegraphics[width=0.98\textwidth]{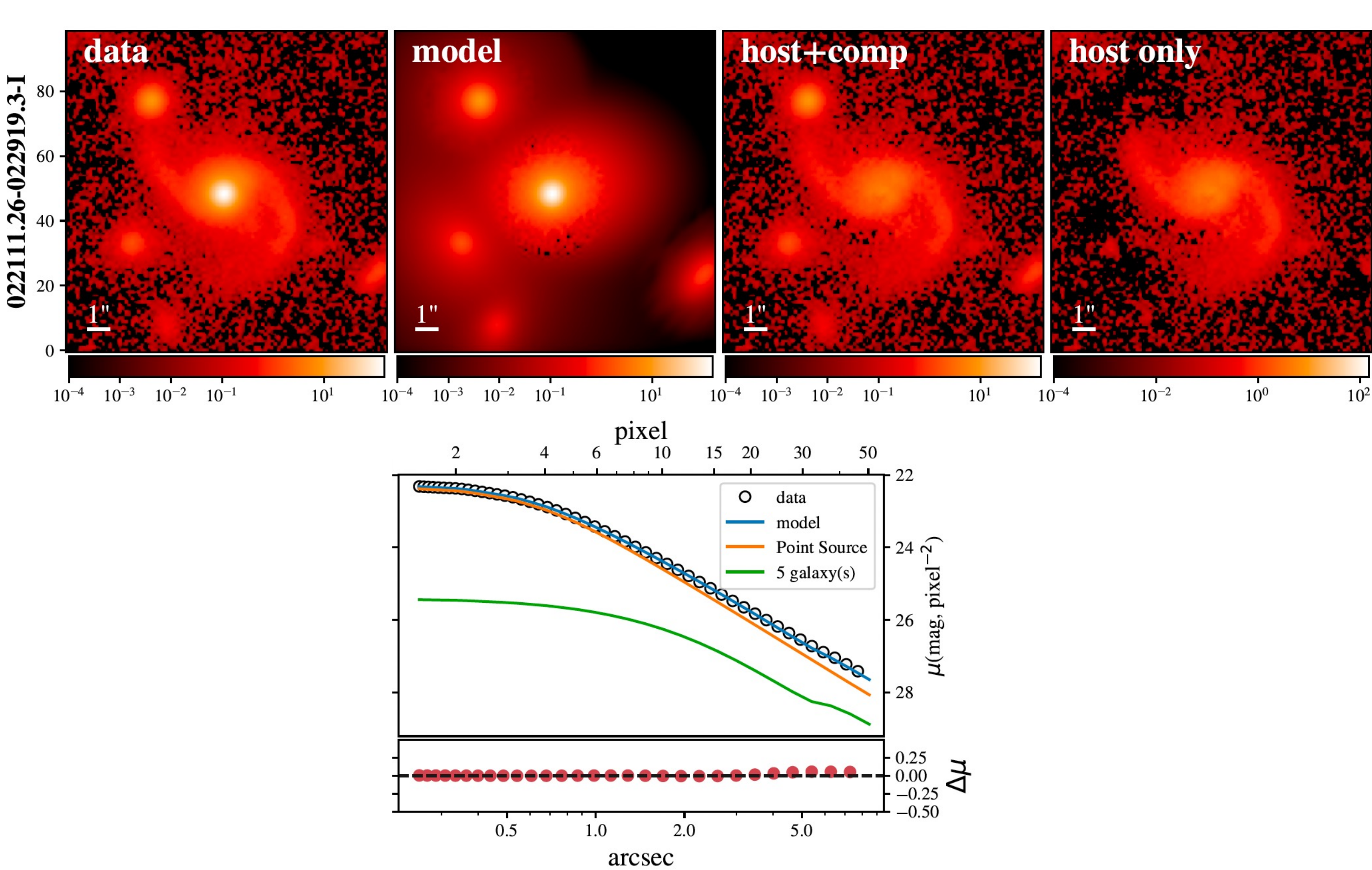}
\end{centering}
\caption{Example 2D decomposition of SDSS J022111.26-022919.3. Top panels from left to right are (1) raw $i$-band HSC image of the quasar; (2) best-fit \textsc{GaLight} model; (3) data minus point source model, we refer to it as ``host+comp" frame; (4) data minus point source minus all the other companions except the quasar host, we refer to it as ``host only" frame. The bottom panel proves the wellness of fitting via comparing the 1D surface brightness profile (as a function of distance from the image center) between model (blue solid line) and data (open circles). The model consists of the point source (orange solid line) and all galaxy components including the quasar host and companions (green solid line). The lower panel presents the residual of data minus model.\label{fig:fit_comp}}
\end{figure*}

To demonstrate the model fitting procedure, we show in Figure \ref{fig:fit_comp} an example (SDSS J000219.11+002232.9) which is a Type-1 quasar hosted by a galaxy with a stellar mass of $10^{10.6} M_{\odot}$ at $z=0.547$. Based on the raw image (data panel) and \textsc{GaLight} modeled image (model panel), we first remove quasar emission, i.e., the point source component, from the raw image. Thus, the quasar host and companions are left, we refer this to the ``host+comp" frame. Because we do not know whether the companions are physically related to the quasar hosts, we make another frame that subtracts the best fit \sersic\ models of all of those companions from the ``host+comp" frame. More importantly, this enables us to assess whether any quasar response to host asymmetries depends on scale from the host to its nearby neighbors. We refer to this frame as the ``host only" frame. We will carry out our measurements on both frames.

\subsection{Parametric quasar host properties} \label{subsec:parametric}
Besides the \sersic\ parameters from \cite{li2021sizes}, we also use their rest-frame U-V color and $M_*$ results estimated by using the SED fitting code CIGALE \citep{boquien2019cigale} based on five-band HSC photometry. They tested that the extrapolation of their SED fitting results to MIR wavelengths agree well with the Wide-field Infrared Survey Explorer (WISE; \cite{wright2010wide}) detections at 3.4 µm after removing the quasar contribution. To qualify the reliability of their results, they selected 1141 CANDELS galaxies in the COSMOS field that are imaged by both HSC and HST. They made mock quasar images via adding PSF models to the HSC images of the CANDELS galaxies to examine how well the decomposition method can recover the real host flux. They reported that the underestimation level is only $\sim$ 7\%. Then they compared HSC- and CANDELS-based measurements on $M_*$ and rest-frame U-V color. They found a small offset of $\sim$ 0.1 dex towards higher $M_*$ in HSC, and the median difference of rest-frame U-V color is around 0.03. They also separated the quasar host types into star-forming galaxies and quiescent galaxies based on a boundary line drawn using the Support Vector Machine algorithm:
\begin{equation} \label{eq:sf_qs}
U-V=0.16 \times \log M_{*}+0.16
\end{equation}
Assuming the CANDELS-based UVJ classification is the ground truth, they estimated that the above classification method can successfully identify 89\% of the massive star-forming hosts with log$M_* > 10.0$ based on HSC photometry. We apply the same UVJ cut to divide the control non-AGN host galaxies. \par

\subsection{Morphological measurements} \label{subsec:morph_measure}
We use \textsc{statmorph}, which is a python-based code developed by \cite{rodriguez2019optical} for estimating both parametric and non-parametric morphological parameters of galaxies. This tool has been tuned based on simulated images from IllustrisTNG and observational data from PanSTARRS at low redshift ($z\sim0.05$). It has been used up to $z\sim4$ \citep{pearson2019effect}. The routine requires a cutout image of the galaxy, a segmentation map that indicates the pixels assigned to the galaxy of interest, and a mask image to reject pixels with most of the flux from other sources. For each of our ``host+comp" and ``host only" images, we generate the segmentation maps and mask images (Section \ref{subsec: MD}) that are then input to \textsc{statmorph} to calculate the asymmetry values (Section \ref{subsec: CAS}).

\begin{figure*}
\begin{centering}
\includegraphics[width=1\textwidth]{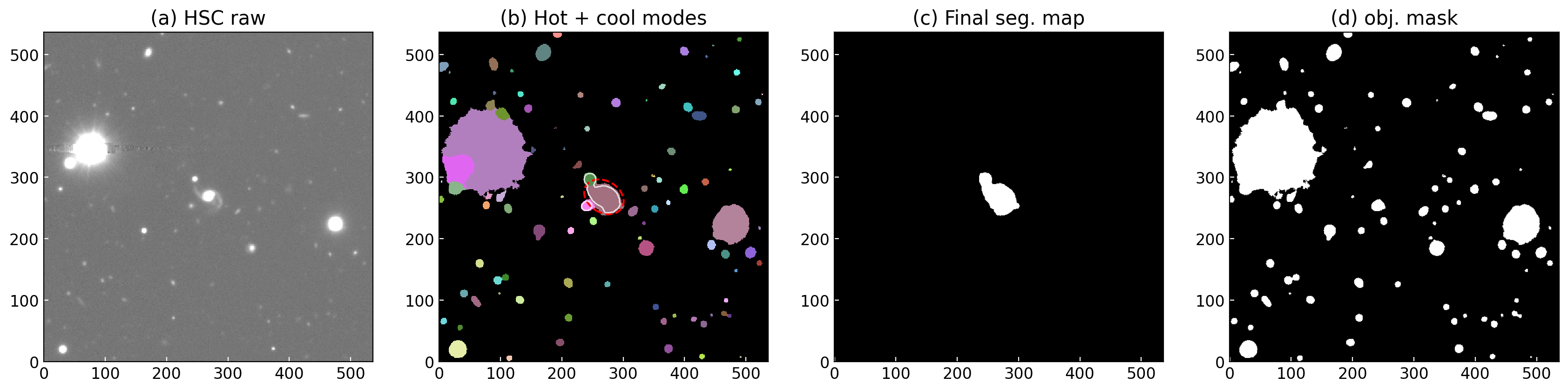}
\end{centering}
\caption{Example production of the segmentation map and mask image (see text for full details) for SDSS J022111.26-022919.3. (a) HSC $i$-band image. (b) Combination of a low \textsc{nsigma} ``cool" mode detection (colored segments) and high \textsc{nsigma} ``hot" mode detection.
(c) All remaining ``cool" segments from the previous step are combined to make the final segmentation map. (d) The other ``cool" segments are combined to create the mask file. \label{fig:segm}}
\end{figure*}

\begin{figure*}
\begin{centering}
\includegraphics[width=0.8\textwidth]{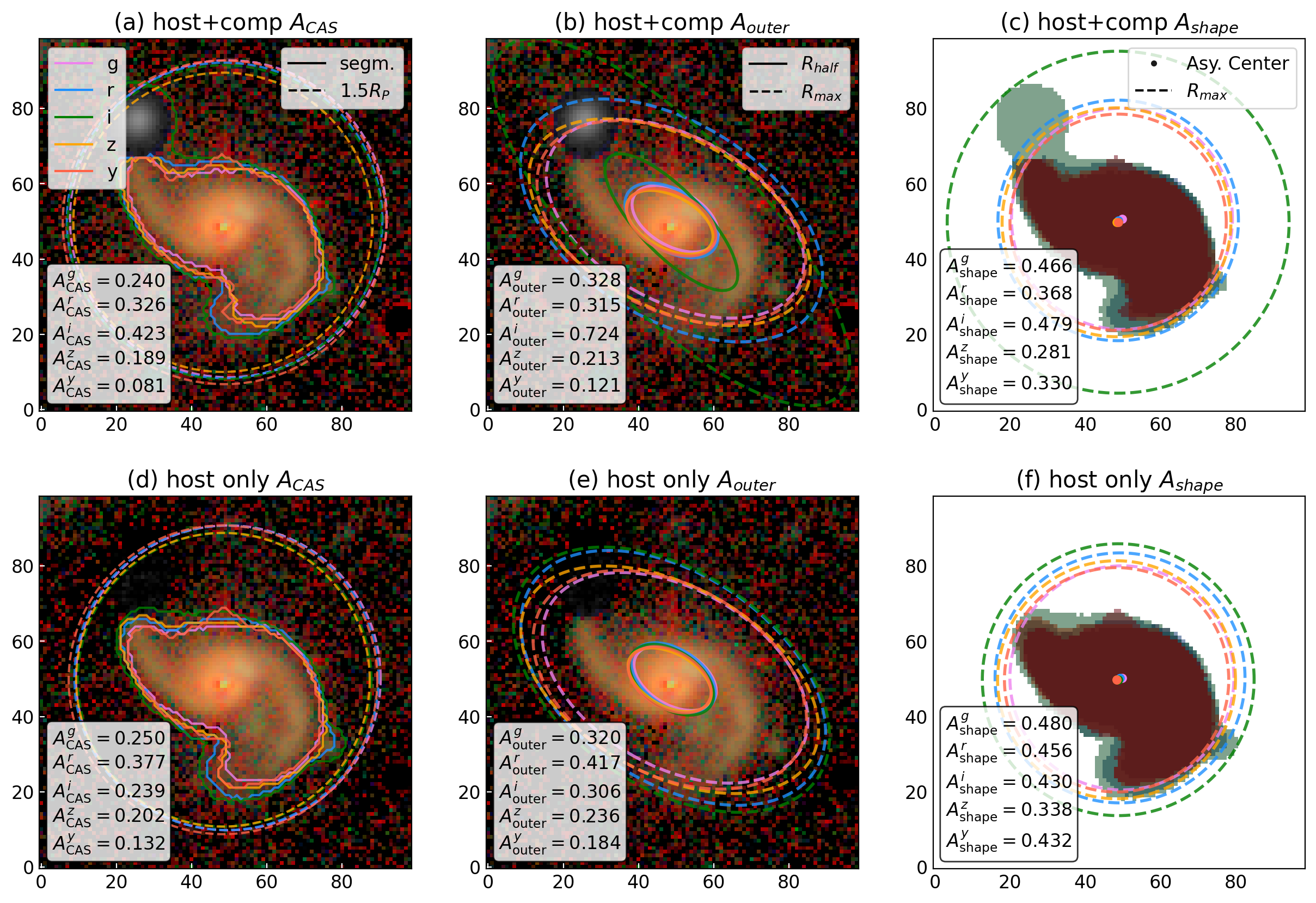}
\caption{Asymmetry measurements of one typical quasar: SDSS J022111.26-022919.3, which is located at $z=0.528$ and with stellar mass of $10^{11.1} M_{\odot}$. For every quasar, we make two image frames: 1st row for the ``host+comp" frame, and the 2nd row for the ``host only" frame. We measure three asymmetry indices for each: $A_{\rm CAS}$, $A_{\rm outer}$, and $A_{\rm shape}$ for each column. Colored images are shown based on all five broad-band HSC data. For $A_{\rm CAS}$ column (panels a,d), we plot the segmentation map envelope with solid lines and 1.5$R_P$ aperture with dashed circles. For $A_{\rm outer}$ column (panels b,e), we plot the half-light radius $R_{\rm half}$ apertures with solid ellipses, and maximum radius $R_{\rm max}$ apertures with dashed ellipses. For $A_{\rm shape}$ column (panels c,f), we show the segmentation map of the object for all five bands, overplotted with the asymmetry center (dots), and $R_{\rm max}$ apertures in dashed circles. For all panels, colors of the contours refer to the bands as labeled in panel (a), from which 5 asymmetry values are measured for every panel.\label{fig:QA}}
\end{centering}
\end{figure*}

\subsubsection{Source deblending and masking} \label{subsec: MD}

We use SExtractor \citep{Bertin1996} for source extraction and deblending that is available in python through the \textsc{photutils} package \citep{larry_bradley_2022_6825092}. The three main parameters of concern are as follow. \par

\begin{enumerate}
 \item \textsc{nsigma}: Detection threshold for a single pixel in units of the standard deviation of the background noise level.
 \item \textsc{npixels}: Minimum number of connected pixels required to be considered a single source
 \item \textsc{contrast}: The flux fraction to deblend a local peak as a separate object from another.
\end{enumerate}



Other parameters are set to default values. However, an optimal separation of the target quasar and close companions often requires a case-by-case tuning of these parameters which is not feasible for a large sample. \cite{sazonova2021all} followed the work of \cite{galametz2013candels} and implemented a so-called ``hot+cool" method that can be effectively applied to a large and diverse set of galaxy images. The basic idea of this method is to use a low \textsc{nsigma} "cool mode" detection to include faint structures such as tidal tails and then apply a high \textsc{nsigma} "hot mode" detection to remove nearby contaminating sources. In Figure \ref{fig:segm}, we demonstrate our implementation of this ``hot+cool" method using the $i$-band image of SDSS J022111.26-022919.3, an SDSS quasar at $z=0.53$. The steps are described in detail as follows:\par

\textbf{Step 1: ``cool" mode.} Using the HSC science image (Figure \ref{fig:segm}a), we first perform a ``cool" mode detection with a $1\sigma$ threshold, 5 pixels minimum area, and $1e-6$ contrast fraction (the same as \cite{sazonova2021all}) to find all faint sources within the field of view, as shown by the colored segments in Figure \ref{fig:segm}b.\par

\textbf{Step 2: Masking unassociated neighbors.} We identify the central source and use its kron radius $R_k$ \citep{kron1980photometry} given in the \textsc{SourceCatalog} as determined by \textsc{photutils}. We then make an elliptical aperture using $5\times R_k$ to include the faint features of the galaxy as shown by the red dashed ellipse in Figure \ref{fig:segm}b. We mask the detected objects outside $5\times R_k$. However, we retain pixels for nearby sources if they have a cool segment with at least $0.01\times A_k$ (area within kron aperture) pixels within the aperture of the main target. As demonstrated in Figure \ref{fig:segm}b, two faint components (green and pink connected segments) pass this step.\par 

\textbf{Step 3: ``hot" mode.} We then apply a ``hot" mode detection routine using a $2.5\sigma$ threshold and no deblending. The segments resulting from this ``hot" mode are plotted as white contours in Figure \ref{fig:segm}b. To save computational time, the ``hot" mode detection is only applied to sources left by Step 2.\par

\textbf{Step 4: Masking close contaminants.} The purpose of the ``hot" mode detection is to check the deblending results based on the ``cool" mode, which may inadvertently split a galaxy into multiple components. Specifically, for each “cool” segment (colored regions), if it covers over 80\% of any “hot” segment (white contours), we consider it as a separated bright peak that is then masked. In Figure \ref{fig:segm}b, the pink ``cool" segment completely covers its ``hot" segment thus determined to be a contaminant. While the ``hot" segment counterpart of the green ``cool" segment covers only a fraction of the large ``hot" segment thus it's considered an important (i.e., possibly interacting) component of the central source. Also, if the area of a ``hot" segment more than doubles that of the central galaxy, we mask its ``cool" segment. Considering the brightness and redshift of our Type-1 quasars, they are unlikely to be secondary members of a minor merger. Such large ``hot" segments are more likely to be foreground galaxies or saturated stars.\par

\textbf{Step 5: Generate final segmentation map and object mask.} Finally, we combine the ``cool" segments from Step 4 to produce the final segmentation map (Figure \ref{fig:segm}c), and all the other ``cool" segments as objects to be masked (Figure \ref{fig:segm}d)\par

We apply this procedure to the ``host+comp" images of our Type-1 quasars. The colored solid contours in Figure \ref{fig:QA}a denote the final segmentation map output from the pipeline. For the ``host only" images, we only perform a simple ``cool" mode detection with $1\sigma$ threshold and $5$ pixels minimum area, without any deblending, to include everything remaining in the frame (Figure \ref{fig:QA}d). This is because all neighbors have been removed during the image-fitting procedure described in Section \ref{subsec:decomposition}.\par

\subsubsection{Asymmetry measurements} \label{subsec: CAS}

The CAS (concentration, asymmetry, smoothness) system is a commonly used non-parametric means to characterize the structure of galaxies \citep{conselice2014evolution}. In this work, we focus on the asymmetry values with three definitions: $A_{\rm CAS}$, $A_{\rm outer}$, and $A_{\rm shape}$. All can be generally written in the following form as defined in \cite{conselice2000asymmetry}:
\begin{equation} \label{eq:A}
A=\min \left(\frac{\Sigma\left|I_{0}-I_{180}\right|}{\Sigma\left|I_{0}\right|}\right)-\min \left(\frac{\Sigma\left|B_{0}-B_{180}\right|}{\Sigma\left|I_{0}\right|}\right)
\end{equation}
For $A_{\rm CAS}$, given the segmentation map of a source (solid envelopes in Figure \ref{fig:QA} left column), $I_0$ is the image pixel values within 1.5 Petrosian radius $R_P$ ($\eta = 0.2$, \cite{petrosian1976surface}, dashed circles in Figure~\ref{fig:QA} left column). $I_{180}$ is rotated from $I_0$ by $180^{\circ}$, which is then subtracted from $I_0$. The total absolute residual values are normalized by the total values of the original $I_0$. This constitutes the first term, namely, object asymmetry. However, this value could be biased due to underlying background gradients \citep{conselice2000asymmetry}. Therefore, the second term, background asymmetry (we refer to it as $A_{\rm bkg}$) is subtracted from the first term to get the intrinsic object asymmetry. \par

\textsc{statmorph} measures the background asymmetry with a so-called "sky box" method. With a maximum box size of 32$\times$ 32 pixels, it iterates over the whole image to find an empty area that includes no pixels from detected objects. If there is no such area with the maximum size, the length of the box side is reduced by a factor of 2 and iterates until an empty area is found. 
We measure the sky asymmetry on the images of the quasars with the size of $90 \arcsec \times 90 \arcsec$ before any model subtractions or cutout using \textsc{GaLight}. This value is then applied to our measurements on "host+comps" and ``host only" frames. Further proof and refinement of the sky asymmetry measurements are described in Appendix \ref{sec:refinements}. \par

As a caveat on the traditional $A_{\rm CAS}$ measurement for bulge-dominated galaxies, the asymmetry index could be underestimated because $R_P$ is effectively smaller thus causing the outskirt features to be ignored. To compensate for this, \citet{wen2014probing} suggested a new asymmetry index, namely the outer asymmetry $A_{\rm outer}$ that is calculated in a similar way to $A_{\rm CAS}$. It replaces the 1.5$R_P$ circular aperture with an elliptical annulus from the half-light radius ($R_{\rm half}$, solid ellipse in Figure \ref{fig:QA}b) that contains half of the total flux within $R_{\rm max}$ and a maximum radius of the target emission ($R_{\rm max}$, dashed ellipse in Figure \ref{fig:QA}b). The elongation and orientation of the ellipse are calculated based on the segmented source image. By doing so, the central region of the galaxy is masked, and the asymmetry value will be more sensitive to the outskirt of the galaxy. \par

As another modification, \cite{pawlik2016shape} defined the shape asymmetry ($A_{\rm shape}$) to increase the sensitivity to faint features by converting all pixel fluxes of the galaxy to a binary mask (Figure \ref{fig:QA}cfi shadowed regions, all five bands are overlapping). Thus, $A_{\rm shape}$ carries information only about the shape of the segmentation map, as shown in Figure \ref{fig:QA}c. It is calculated within a circular aperture with radius $R_{\rm max}$ (Figure \ref{fig:QA}cfi dashed circle), and no background asymmetry subtraction is required. \par

In total, we calculate 30 asymmetry values for every Type-1 quasar, combining five photometric bands, two image frames (``host+comp" and ``host only"), and three asymmetry indices.



\begin{figure}
\begin{centering}
\includegraphics[width=0.45\textwidth]{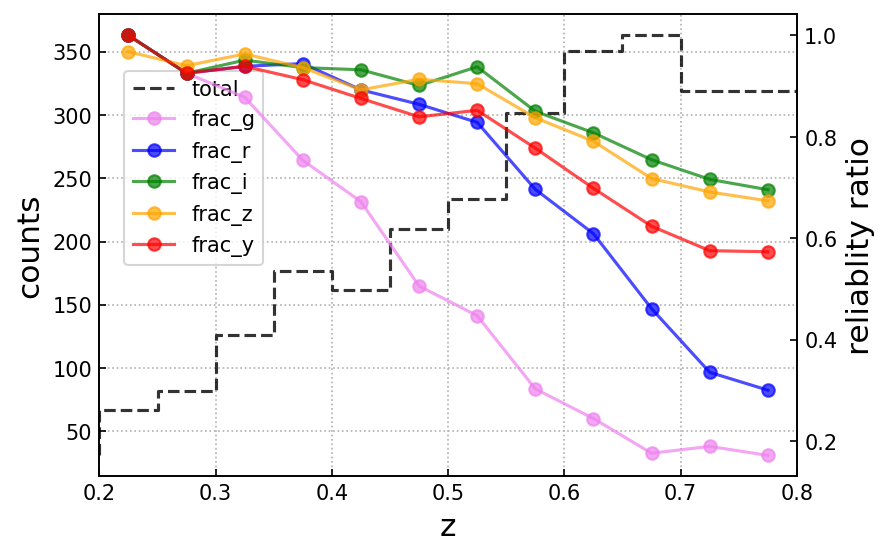}
\caption{Ratio of our asymmetry measurements as a function of redshift which indicates the level of reliability. A target being assessed with \textsc{statmorph} $flag = 0$ and successfully passing the entire mask and deblending pipeline are considered a reliable measurement. The dashed histograms are the total number of objects in every redshift bin (left y-axis), and the corresponding ratios are shown in colored broken lines for each band (right y-axis). 
\label{fig:completeness_Q}}
\end{centering}
\end{figure}

\begin{figure*}
\begin{centering}
\includegraphics[width=0.9\textwidth]{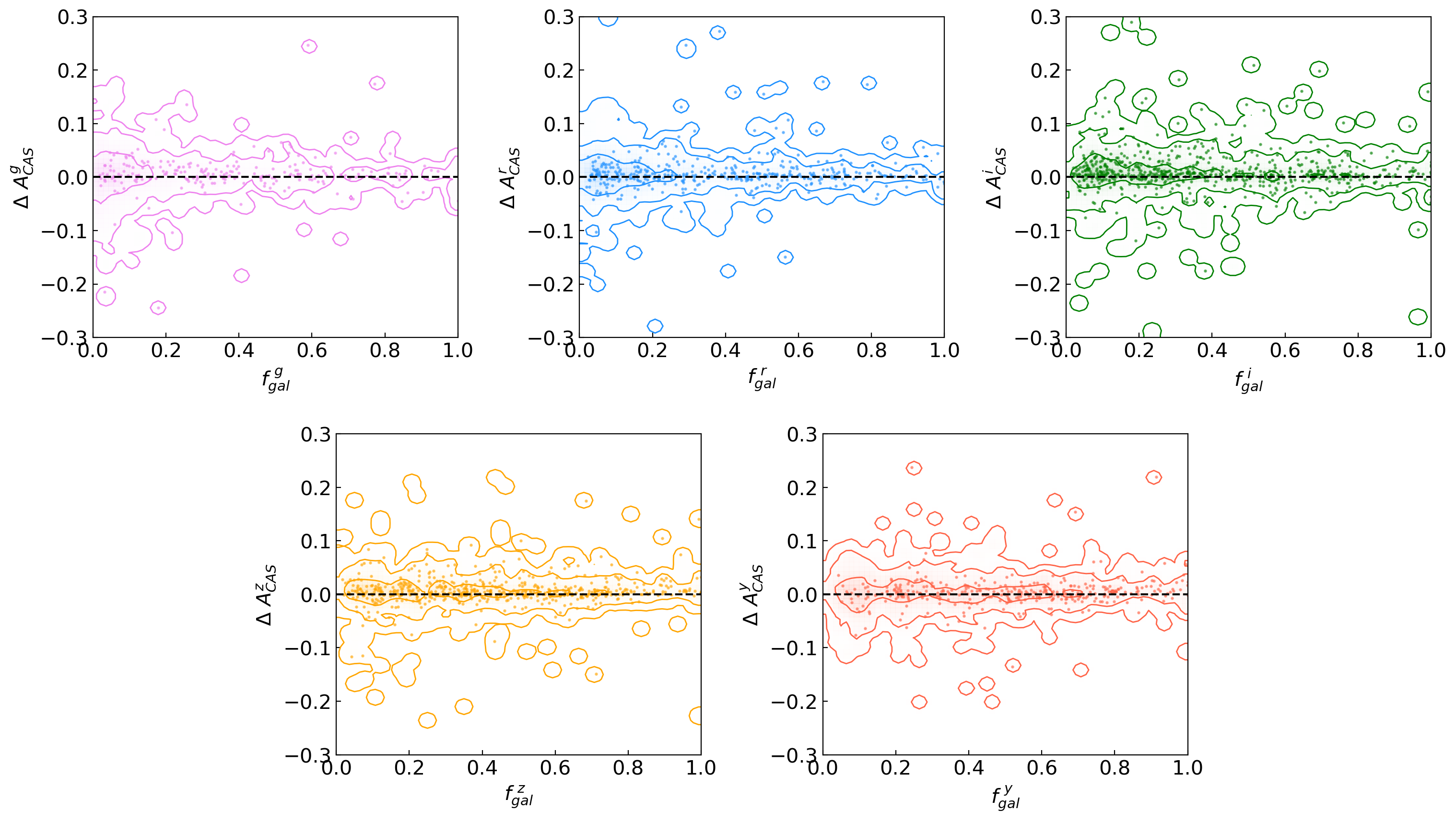}
\end{centering}
\caption{\textcolor{black}{Quality assessment (i.e., asymmetry recovery tests) measured as an offset ($\Delta$) of $A_{\rm CAS}$ between asymmetry values before and after decomposing the simulated quasar images in each of the five bands separately. The x-axis refers to the host-to-total flux ratio ($f_{\rm gal}$). The contours show the entire sample, which are $1\sigma$, $2\sigma$, and $3\sigma$ envelopes of binned 2d histogram distributions (i.e., 99.7\% data points are located within the outermost contours). Similar contours in plots hereafter will use the same definition. The colored dots show the subset after applying \textsc{statmorph} flags. The dashed black lines mark the zero offsets. We find the flags can effectively reduce the cases that might have quasar subtraction issues at low $f_{\rm gal}$.}\label{fig:deltaA_candels}}
\end{figure*}

\begin{figure}
\begin{centering}
\includegraphics[width=0.48\textwidth]{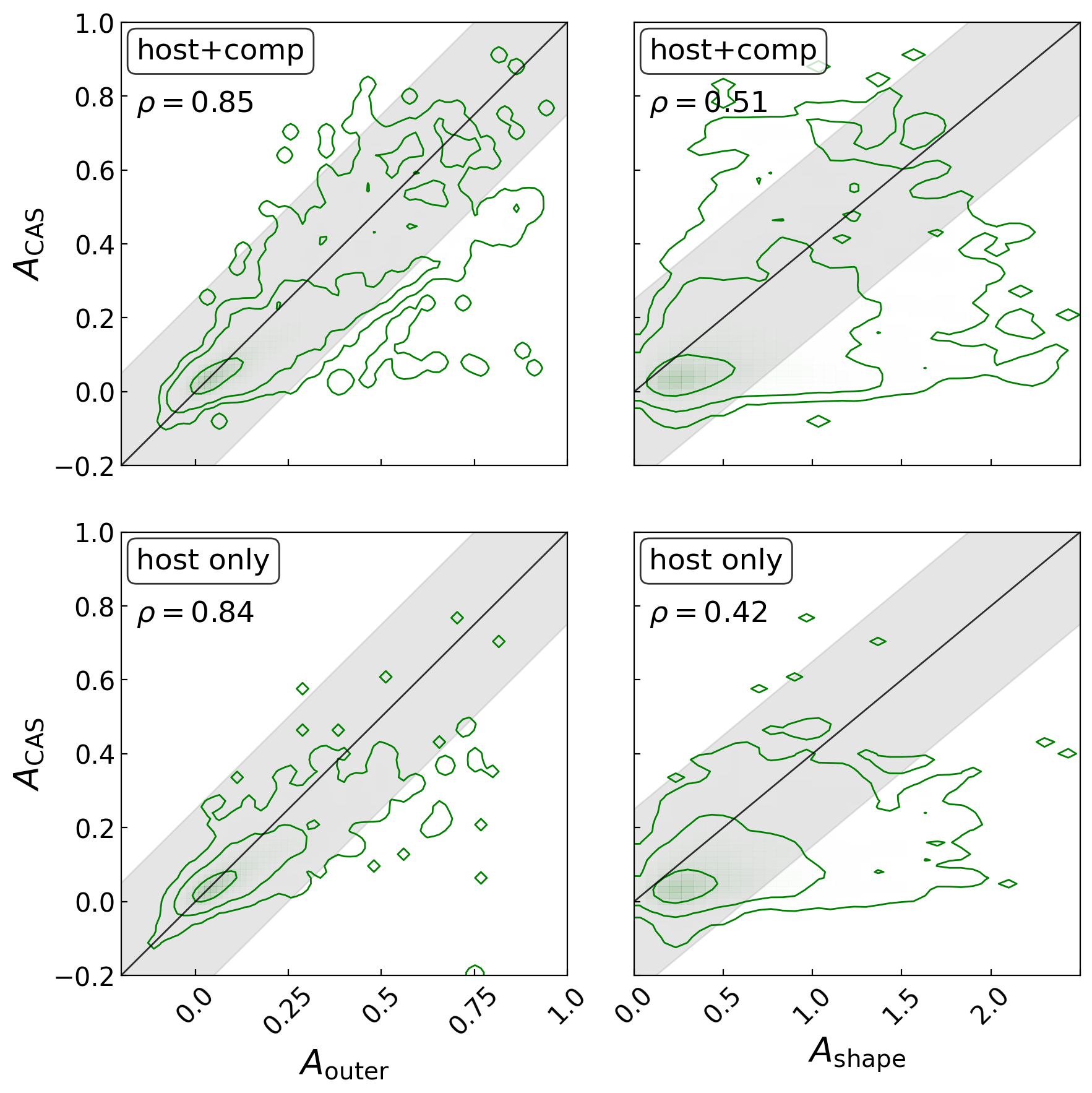}
\caption{\textcolor{black}{Comparisons between $A_{\rm CAS}$, $A_{\rm outer}$ (left column), and $A_{\rm shape}$ (right column), for ``host+comp" frames (top row) and ``host only" frames (bottom row) in the $i$-band. } The black solid lines indicate a 1:1 relation, and the grey shadowed regions indicate $\pm 0.25$ offsets from that. We give the Pearson correlation coefficient $\rho$ in the upper left corner of each panel. \label{fig:qualify_Q}}
\end{centering}
\end{figure}

\begin{figure*}
\begin{centering}
\includegraphics[width=0.85\textwidth]{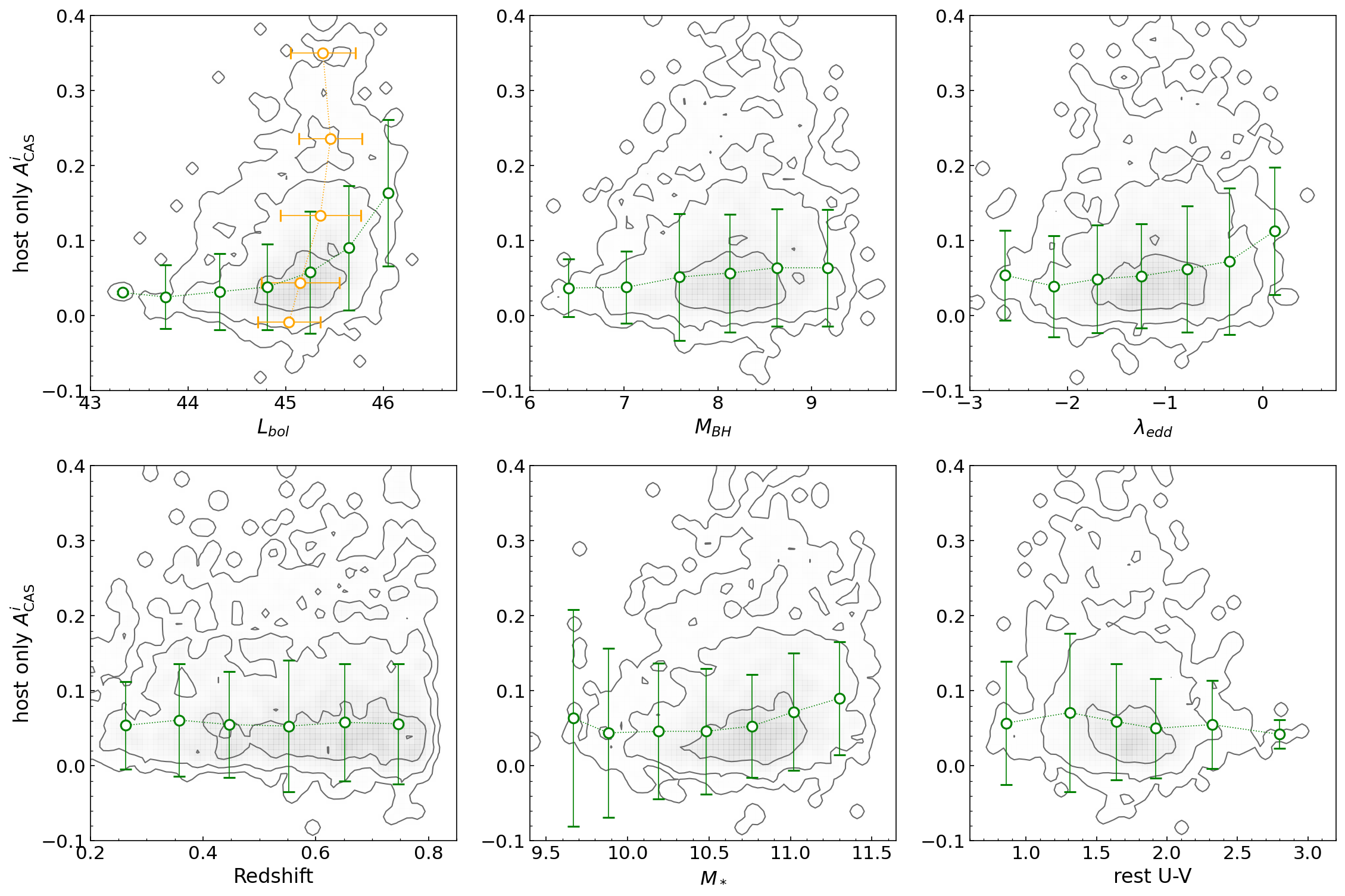}
\end{centering}
\caption{$A_{\rm CAS}^i$ as a function of quasar and host galaxy properties. \textcolor{black}{Contours show 1, 2 and $3\sigma$ envelopes of the 2D distributions. Median values are based on binning either according to their physical property (shown on the x-axis, green open circles) or the $A_{\rm CAS}^i$ values (y-axis, orange open circles). Error bars represent the standard deviation of each binned subset. We highlight the top-left panel that shows the strongest correlation with asymmetry and has an upturn at $\log~L_{\rm bol}=45$.}  \label{fig:host_A_evolution_single}}
\end{figure*}

\subsection{Quality assessment} \label{subsec:qualification}
We assess the reliability of our asymmetry measurements for both the ``host+comp" and ``host only" science frames separately. Reasons for inaccurate measurements could be multi-fold which originate mainly from the \textsc{statmorph} flags \cite{rodriguez2019optical}, and a few from the mask and deblending pipeline (Section \ref{subsec: MD}, minor). Issues can include quasar hosts undetected in ``cool" mode, residuals of the subtraction may lead to an improper Petrosian radius, or the background not being properly assessed. \textcolor{black}{We show the ratio of the measurements that are free of these issues as a function of redshift in Figure \ref{fig:completeness_Q}, as a means to gauge the level of reliability in the asymmetry measurements.} The dashed histograms indicate the total number of objects in each redshift bin, and the colored broken lines are the reliability ratios in each band. Generally, the ratio of reliable measurement is highest in $i$-band, while it drops rapidly towards higher redshift and in bluer bands. This is because the $i$-band wavelength coverage (695–845 nm) is less affected by sky background noise and galactic extinction. The survey was designed to take $i$-band images under better seeings and darker nights \citep{aihara2017hyper}. These factors make the quality of $i$-band images better than the other four, thus making the reliable ratio of measurements higher in this band. In total, the numbers of targets with reliable measurements for the five bands are: $g$, 886; $r$, 1488; $i$, 1987; $z$, 1939; $y$, 1759, respectively. We also checked the dependence of the reliability ratio on other properties, we find that it is almost independent of $M_{\rm BH}$ and $\lambda_{\rm edd}$ (the difference between each bin is within 0.1 in $i$ band). While the ratio is higher when $M_*$ is larger, as the host galaxy is easier to be detected, the difference between the most and least massive bins is $\sim 0.2$. \par
\textcolor{black}{We then use simulated quasar images to assess a possible over-subtraction issue of quasar hosts and its impact on the asymmetry measurements. \cite{li2021sizes} created a set of 1141 simulated quasar images by inserting a model HSC PSF (mock quasar) into HSC images of real inactive CANDELS galaxies matched to the redshift and stellar mass of the quasar hosts. The flux of the mock quasar is set to the actual $i$-band value for the matched HSC quasar. It is then scaled to other bands using a power law function in wavelength. By comparing the original galaxy magnitudes and those after adding and subtracting the mock quasar, they found that the decomposition process results in only a $\sim 7\%$ over-subtraction affecting the host-to-total flux ratio ($f_{\rm gal}$). We reran our asymmetry measurement pipeline on both the original and processed galaxy images and compared the results in terms of offsets of in $A_{\rm CAS}$ ($\Delta A_{\rm CAS} = A_{\rm original} - A_{\rm processed}$). As shown by the contours in Fig. \ref{fig:deltaA_candels}, the entire sample has $1\sigma$ scatter of $\sim 0.05$ for $\Delta A_{\rm CAS}$. There is no strong bias on whether the subtraction leads to systematically larger or smaller asymmetry values. Even so, the data are more scattered at low $f_{\rm gal}$, i.e., luminous quasars. The impact of the over-subtraction is stronger in bluer bands as the quasars are brighter in those bands. We then apply the \textsc{statmorph} flags and only keep the reliable measurements (colored dots in Fig. \ref{fig:deltaA_candels}) that result in the scatter being significantly reduced. Taking the $i$-band as an example, if we define $|\Delta A_{\rm CAS}|>0.1$ as a significant offset, then only $\sim 4.5\%$ of the sources are affected by subtraction issues, and $\sim 50\%$ of those are flagged out by \textsc{statmorph}. This is because the strange residuals of bad-subtracted cases would more likely lead to a failure of \textsc{statmorph}, e.g., in finding a proper asymmetry center. }\textcolor{black}{Therefore, we suggest that for our reliably measured quasar hosts, only $\sim 2\%$ are affected by significant subtraction issues. For the following main results of this work, we only use these reliable measurements (a test of keeping the statmorph-rejected sources show consistent results, see Appendix \ref{subsec:uncertainties}).} \par

For the reliable measurements, we further compare different asymmetry indices for both the ``host only" and ``host+comp" frames in Figure \ref{fig:qualify_Q}. The left column compares $A_{\rm CAS}$ with $A_{\rm outer}$, and right column compares $A_{\rm CAS}$ with $A_{\rm shape}$. The $i$-band data are shown in $1\sigma$, $2\sigma$, and $3\sigma$ contours. The black solid lines indicate 1:1 relations in each panel. Grey shadowed regions indicate $\pm 0.25$ offsets from the 1:1 relations. We show the Pearson correlation coefficient $\rho$ between the asymmetry indices at the left upper side of each panel. For all bands and both frames, $\rho$ between $A_{\rm CAS}$ and $A_{\rm outer}$ is $0.84-0.9$, and is $0.37-0.56$ between $A_{\rm CAS}$ and $A_{\rm shape}$. Compared to ``host+comp" measurements, the removal of companion sources generally reduces all asymmetry values, making $A_{\rm CAS}$ and $A_{\rm outer}$ values more constrained to be below 0.5 in ``host only" measurements. $A_{\rm shape}$ has larger scatter than $A_{\rm outer}$, but still has a moderate correlation with $A_{\rm CAS}$.\par


\section{Results} \label{sec:results}
\subsection{Correlations between asymmetry and quasar properties} \label{subsec:Quasar_A_corr}

\textcolor{black}{We first consider the dependence of asymmetry on quasar and host properties including} $L_{\rm bol}$, $M_{\rm BH}$, $\lambda_{\rm edd}$, redshift, stellar mass $M_*$, rest-frame U-V color. We show the distribution of $A_{\rm CAS}^i$ measured on ``host only" frames as a function of these properties in Figure \ref{fig:host_A_evolution_single}. \textcolor{black}{Here, the use of ``host only" frames reduces the influence of projection effects (i.e., foreground and background objects) on the asymmetry measurements. We bin the measurements according to their physical properties with fixed intervals (0.5 dex for $L_{\rm bol}$, $M_{\rm BH}$, and $\lambda_{\rm edd}$; 0.3 dex for $M_*$; 0.4 mag for rest-frame U-V; 0.1 for redshift)}. \textcolor{black}{For each bin, median values are shown as green open circles with standard deviation represented by the error bars}. We find that $A_{\rm CAS}^i$ has moderate positive correlations with $L_{\rm bol}$, $M_{\rm BH}$, $\lambda_{\rm edd}$, $M_*$, and a weak negative correlation with rest-frame U-V color of the host galaxy. There is no clear correlation with redshift. While not shown, we also tested with the observed host galaxy color ($g-r$); it shows a weak negative correlation with $A_{\rm CAS}^i$, similar to the rest frame $U-V$ color. We highlight the $L_{\rm bol}$-A correlation turns out to be the strongest among these parameters, which will be discussed in more detail in the following sections. Here, we first point out a turning point at $L_{\rm bol}\sim45$, beyond which the correlation appears to strengthen.

\textcolor{black}{In addition, we bin the data by $A_{\rm CAS}^i$ (in intervals of 0.1, orange open circles in the top left panel. 
$A_{\rm CAS}$ is likely related to longer term activity of a quasar host galaxy, thus binning by $A_{\rm CAS}$ might smooth out short term variations. In this way, we find that more asymmetric hosts have slightly more luminous quasar activity. This is similar to what we find by binning along the x-axis values.} \par


\begin{figure}
\begin{centering}
\includegraphics[width=0.45\textwidth]{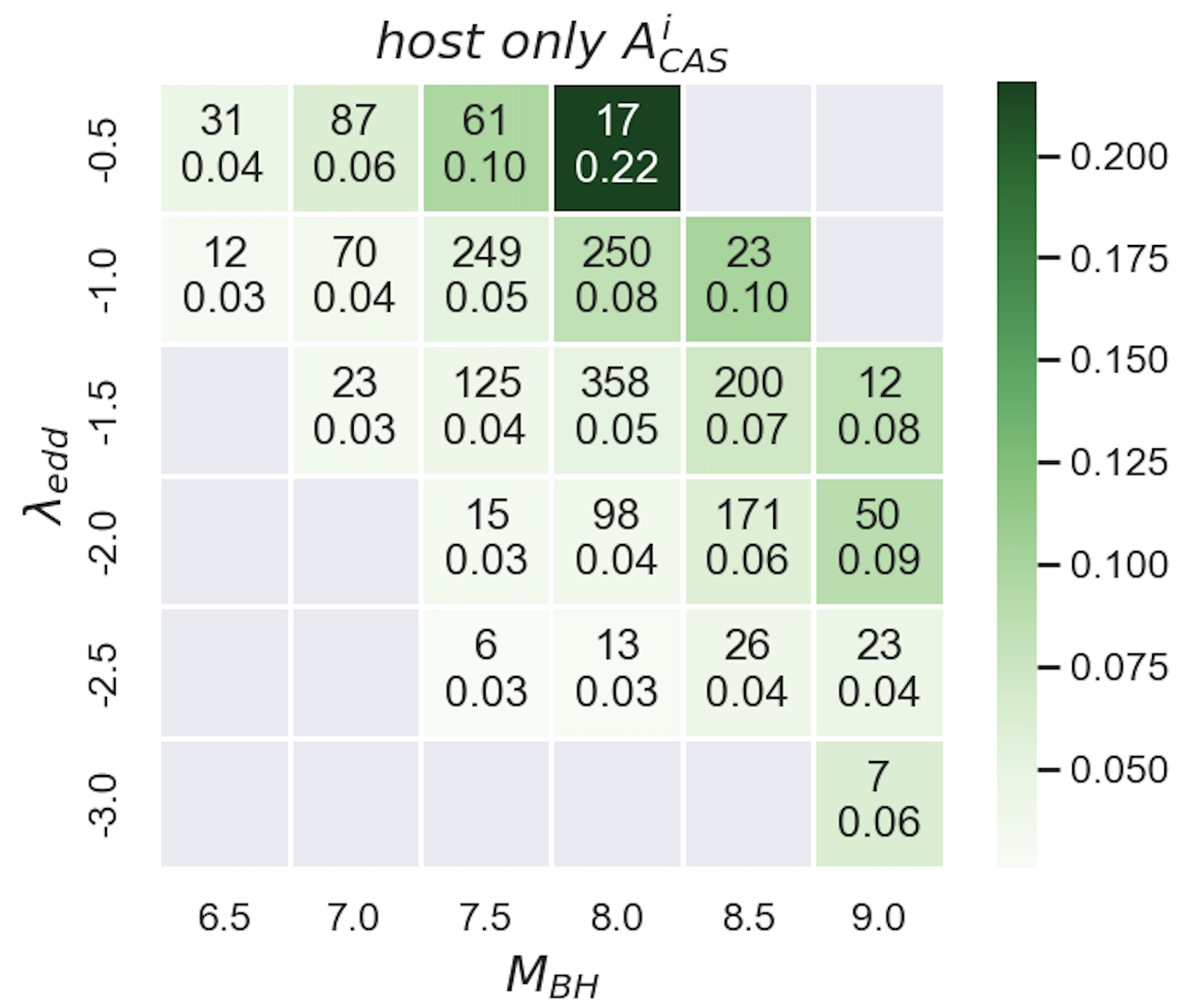}
\end{centering}
\caption{\textcolor{black}{Median} asymmetry of quasar hosts in bins of BH mass and Eddington ratio. In each cell, the top value shows the number of objects per bin, the lower number shows the median asymmetry value of these objects. Only cells with more than five objects are kept. The value of the asymmetry is indicated by the scale bar on the right. We report a trend from lower left to upper right that when $M_{\rm BH}$ and  $\lambda_{edd}$ increase, the host galaxy of quasars is slightly more asymmetric on average. \label{fig:heat_BH}}
\end{figure}

\begin{figure*}
\begin{centering}
\includegraphics[width=0.8\textwidth]{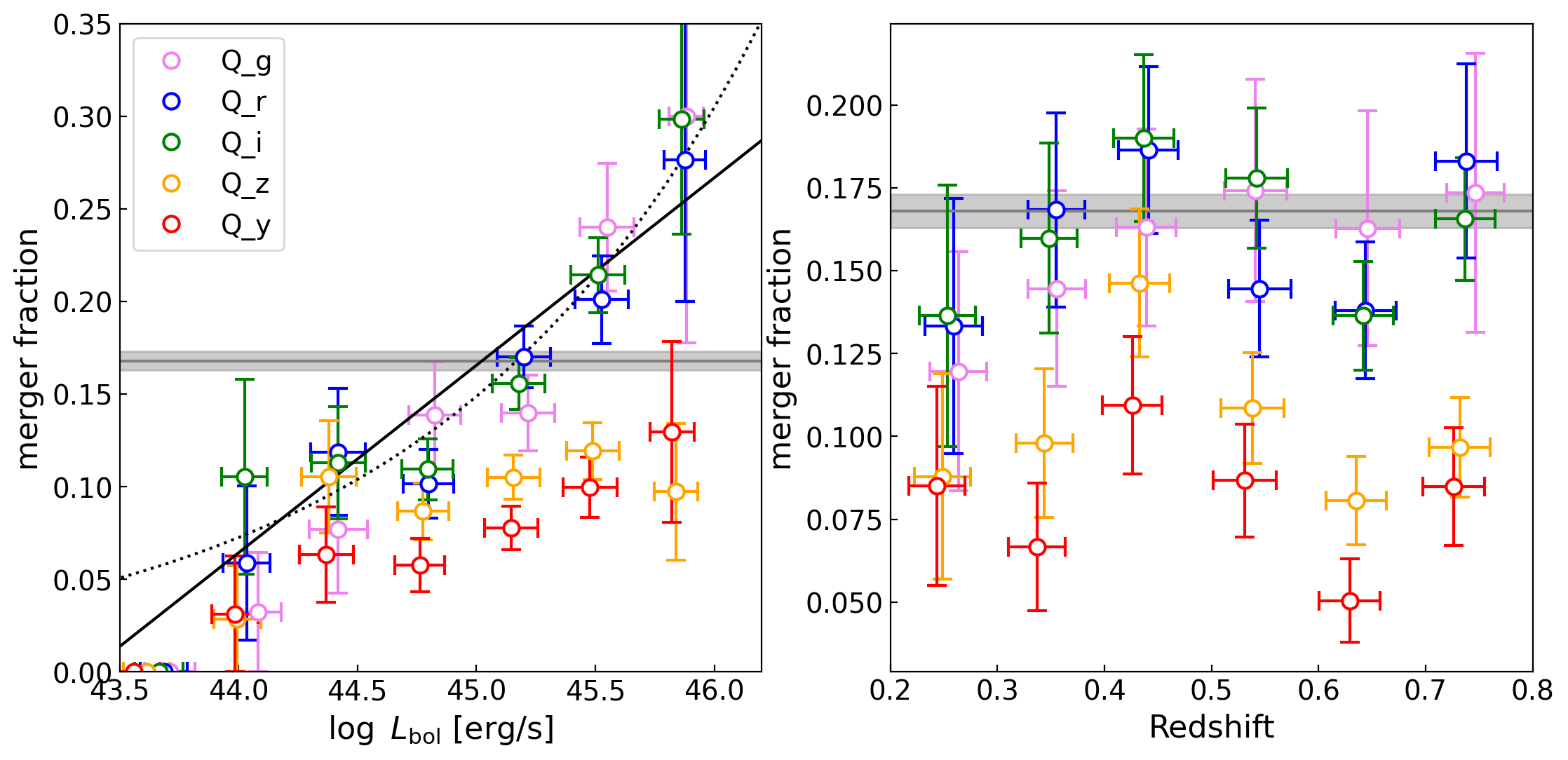}
\caption{Merger fractions of quasar hosts as a function of $L_{\rm bol}$ (left) and redshift (right) in all five bands (Q\_grizy). The error bars in the merger fraction are estimated from the Poisson errors of mergers. \textcolor{black}{The x-values of the points are shifted a bit for clarity of the data in each band. The $i$-band data for quasar hosts in the left panel is fitted with a linear relation (solid line, Equation \ref{eq:linear}) and a power law (dashed line, Equation \ref{eq:power}). The merger fraction for the control galaxy sample in the $i$-band is shown as the grey line with shadowed Poisson error}. \label{fig:merger_fraction}}
\end{centering}
\end{figure*}

\textcolor{black}{We further} quantify the typical asymmetries of our quasar sample by the properties of its BH and host galaxy. We first bin our quasars in $M_{\rm BH}$ from $10^{6.5}$ to $10^{9.5} \mathrm{M_{\odot}}$ and $\lambda_{edd}$ from -3 to 0 both with intervals of 0.5 dex (Figure \ref{fig:heat_BH}). We only keep bins that include more than five objects. The counts are shown by the upper number in each bin. Then we calculated the median asymmetry values as noted by the lower number in each bin. The darker color of the cell indicates a larger median asymmetry value. We see a trend from the lower left to the upper right. Therefore, we report that the quasar hosts are more asymmetric when they harbor a more massive and more actively-accreting BH.\par


\begin{figure*}
\begin{centering}
\includegraphics[width=0.85\textwidth]{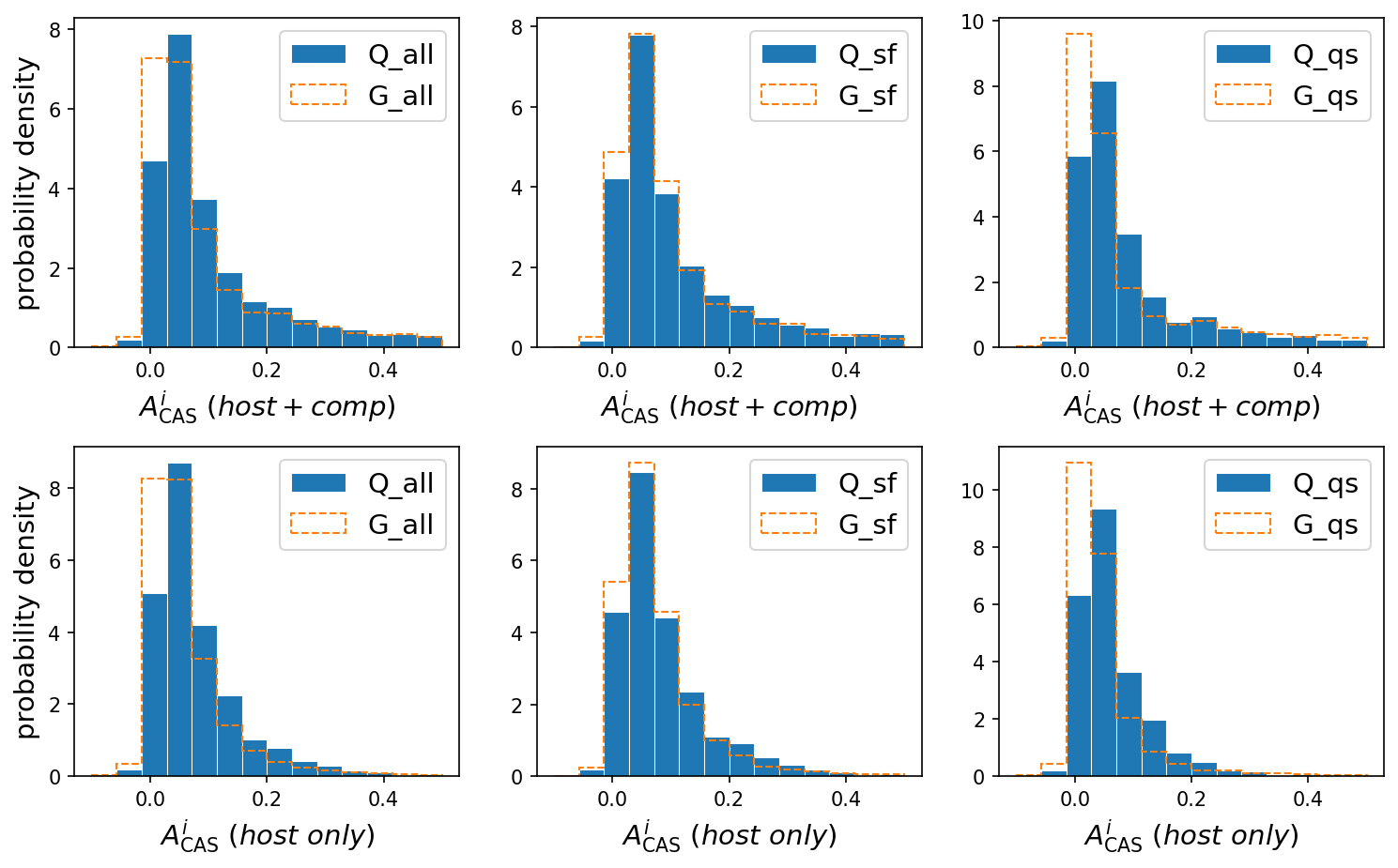}
\end{centering}
\caption{Probability density distribution of $A_{\rm CAS}^i$ of quasar hosts (solid blue histograms) and control galaxies (dashed red histograms). From left to right, the columns show different samples: all hosts, star-forming hosts, and quiescent hosts. The top row is based on the ``host+comp" frames while the bottom row uses the ``host only" frames. \label{fig:QG_A_CAS}}
\end{figure*}

\subsection{Merger fraction} \label{subsec:merger_frac}

\textcolor{black}{We then convert the asymmetry values to merger fractions and study its correlations with $L_{\rm bol}$ and redshift.} Following \cite{conselice2003relationship}, we classify objects with either $A_{\rm CAS}>0.35$ or $A_{\rm outer}>0.35$ as major mergers. We only use ``host+comp" frames for this purpose to minimize the risk of removing genuine companions. However, this estimate is an upper limit because it includes projected nearby contaminants. We show the merger fraction as a function of $L_{\rm bol}$ and redshift in all five bands (Figure \ref{fig:merger_fraction}). \textcolor{black}{In this section, we only focus on the $i$-band results, i.e., the green points. The quasars are binned in a similar way as in Figure \ref{fig:host_A_evolution_single}, but per 0.4 dex for $L_{\rm bol}$ based on the distribution of the samples.} The error bars on the x-axis indicate the standard deviation of $L_{\rm bol}$ and redshift in each bin. The uncertainties on the merger fractions are the assumed Poisson error. 

\textcolor{black}{We find that the merger fractions of quasar hosts are overall low (16.2\% in $i$-band) and have little dependence on redshift (spearman correlation coefficient $\rho=0.2$ with $p$-value=0.7). Therefore, most of the quasars are not triggered by mergers at $z<0.8$.} \textcolor{black}{Following \cite{treister2012major}, we fit a linear relation (black solid line) and a power law (black dashed line) to the $i$-band merger fractions of quasar hosts as functions of $L_{\rm bol}$. The fitting was performed with \textbf{scipy.optimize.curve\_fit} package using least squares optimization.} The best fit result for linear relation is:
\begin{equation}
\operatorname{frac}(L)=\frac{\log \left(L_{\mathrm{bol}}\right)-43.4^{+0.4}_{-0.4}}{9.9^{+2.4}_{-2.4}}
\label{eq:linear}
\end{equation}
and for power law:
\begin{equation}
\operatorname{frac}(L)=\left(\frac{L_{\mathrm{bol}}}{4.51^{+3.4}_{-3.4} \times 10^{47} \mathrm{erg}\  \mathrm{s}^{-1}}\right)^{0.31^{+0.05}_{-0.05}}
\label{eq:power}
\end{equation}

The fitting of \cite{treister2012major} was based on multiple studies, thus could be affected by a lack of uniformity in sample selection and classification criteria. \textcolor{black}{Here our evolution is based on one single sample set and one single criterion of classification.} Nevertheless, the zero point of our fitted linear relation is consistent with \cite{treister2012major} within the error ($43.4\pm0.4$ versus 43.2), which indicates an expected threshold of $L_{\rm bol}$, below which the major merger fraction of AGNs is negligible. Above this threshold, the power law slope is close to that previously reported ($0.31\pm0.05$ versus 0.4), indicating a similar increasing tendency of merger fraction with increasing $L_{\rm bol}$. The difference between the exact values of merger fractions in our work and their work is likely caused by the different classification criteria, which leads to the different scaling of our fitted functions. \par


\subsection{Asymmetry of Quasar hosts and control galaxies} \label{subsec:QG_A_CAS}

We compare the $i$-band CAS asymmetry ($A_{\rm CAS}^i$) of quasar hosts and inactive galaxies in Figure \ref{fig:QG_A_CAS} for both ``host+comp" and ``host only" frames. The left column shows the probability density distribution of $A_{\rm CAS}^i$ for the entire population of quasars and galaxies. The comparisons between star-forming and quiescent types are shown in the middle and right columns. Overall, ``host+comp" frames have longer tails towards higher asymmetries than ``host only" frames in all sub-panels, indicating that the highest asymmetries are due to the companions. Considering the star-forming galaxies only, the median excess of quasar host asymmetry over control galaxies is below 0.01. \textcolor{black}{While for quiescent galaxies, quasar hosts are slight biased toward higher asymmetry. The entire sample of quasar hosts has an excess of 0.017 in median over the control galaxies for the ``host only" frames. The statistical error of the excess (i.e., $\sigma/\sqrt{N}$ of $A_{Q}-A_{G}$) is 0.0018, thus this excess has a significance level of $9.4\sigma$. For the ``host+comp" frames, this excess is $0.014\pm0.004$.}

\begin{figure*}
\begin{centering}
\includegraphics[width=0.9\textwidth]{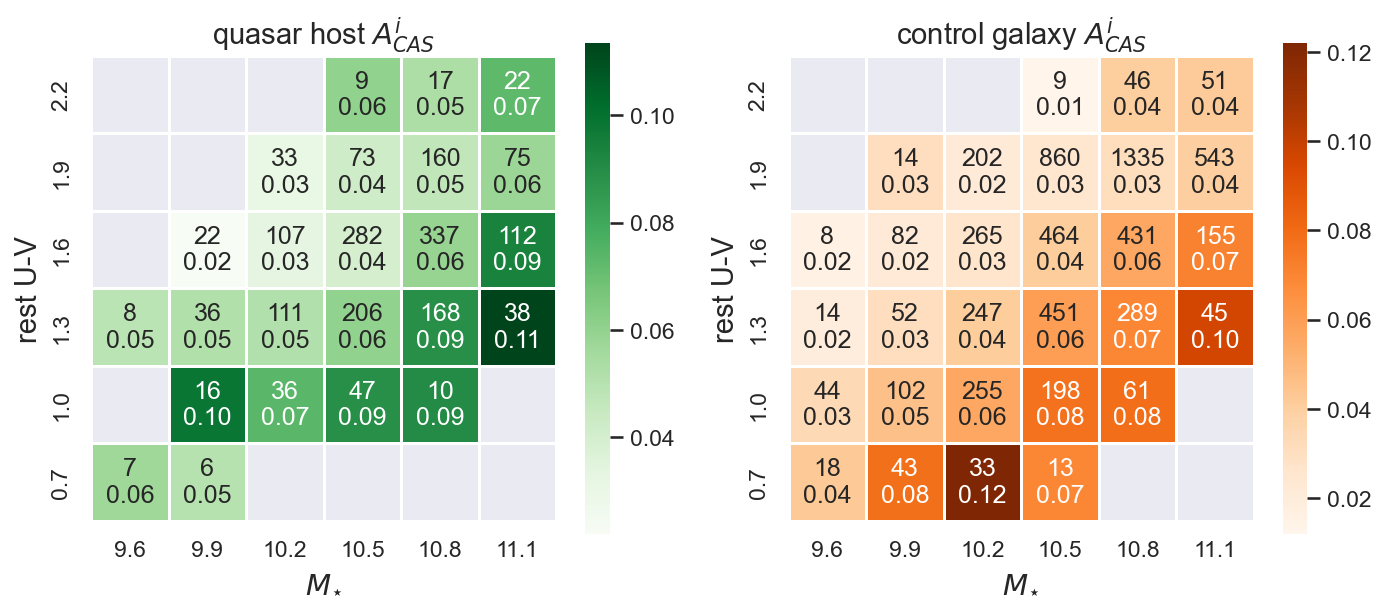}
\caption{\textcolor{black}{Similar to Fig.~\ref{fig:heat_BH}, median asymmetry but now binned with $M_*$ from $10^{9.6}$ to $10^{11.4} M_{\odot}$ per 0.3 dex, and rest-frame U-V host color from 0.7 to 2.5 per 0.3 mag. The left (right) panel is for quasar hosts (control galaxies).
} \label{fig:heat_host}}
\end{centering}
\end{figure*}

Next, we bin our quasars with respect to $M_*$ from $10^{9.6}$ to $10^{11.4} M_{\odot}$ per 0.3 dex, and rest-frame U-V color of the host galaxy from 0.7 to 2.5 per 0.3 mag. \textcolor{black}{We compare these values to that of control galaxies with the same binning. The results, shown in Figure \ref{fig:heat_host}, indicate generally similar trends in both panels that the asymmetry values increase with increasing stellar mass and bluer color. Considering the typical uncertainty level of $A_{\rm CAS}^i$ to be $\sim$0.03 (see Appendix \ref{sec:refinements} for the details on the uncertainties of asymmetry measurements), we find that the presence of a quasar leads to a slight excess of asymmetry as shown in some low mass bins and at the red and high-mass end. A simple interpretation could be that less massive galaxies are more easily affected by AGN feedback which may redistribute gas and hence influence the location of subsequent star formation. Alternatively, if quasars trigger star-forming activity, which lead to asymmetries, it may be a relatively stronger effect in systems that originally had low star-forming activities (quiescent and red galaxies).} \par

\subsection{Band dependencies} \label{subsec:band}

\begin{figure*}
\begin{centering}
\includegraphics[width=0.85\textwidth]{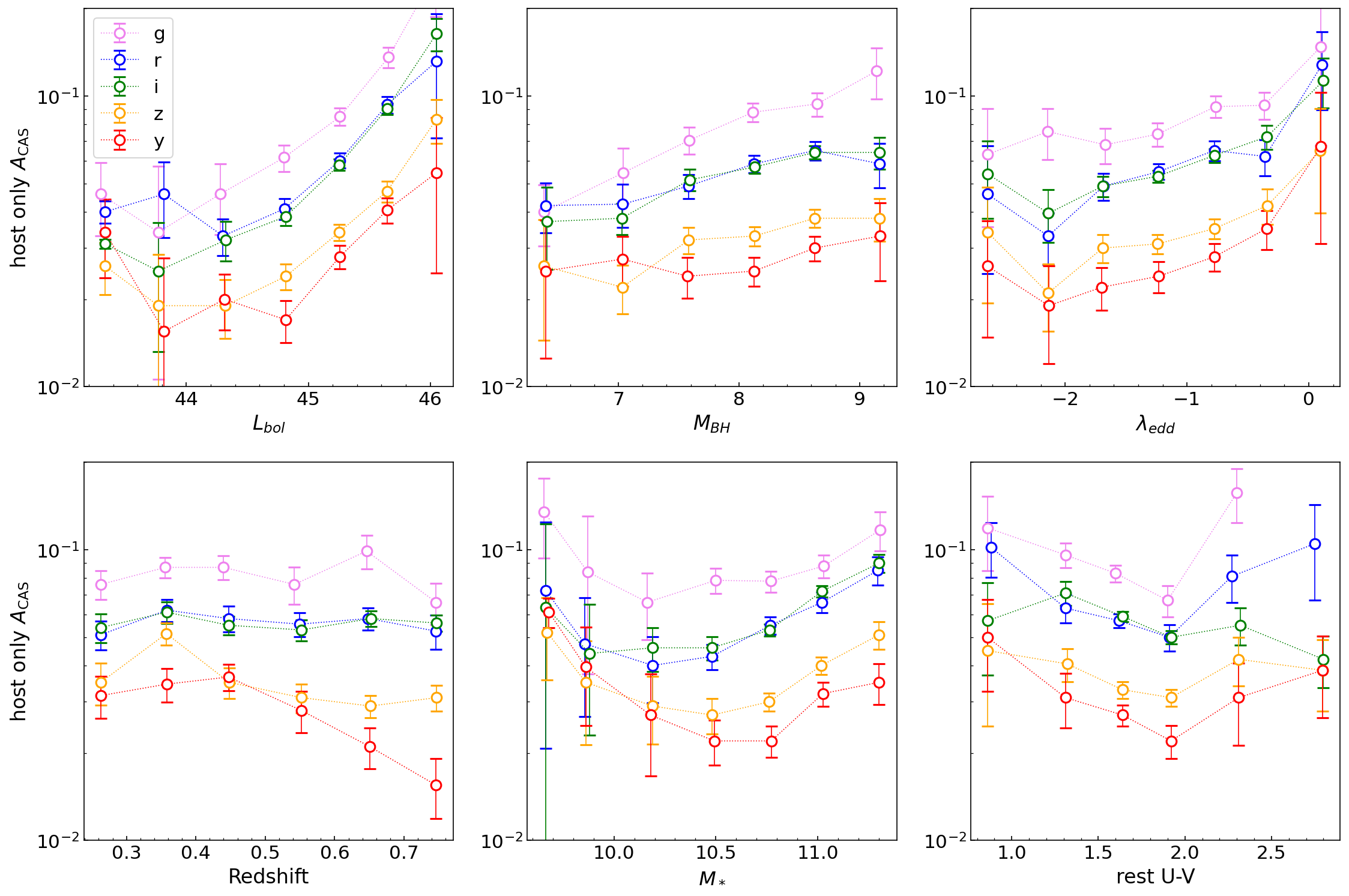}
\end{centering}
\caption{Similar to Fig.~\ref{fig:host_A_evolution_single}, $A_{\rm CAS}^i$ as a function of quasar and host galaxy properties but now shown for all five bands. For clarity, we only show the median values, not all of the individual values, \textcolor{black}{and the y-axis is in log scale}. The vertical bars indicate the error on the mean ($\sigma / \sqrt{N}$) for each bin. \label{fig:host_A_evolution}}
\end{figure*}

As mentioned above, we find that our results depend on the optical band under consideration. We first revisit the correlations, shown in Figure \ref{fig:host_A_evolution_single}, for all five optical bands (Figure \ref{fig:host_A_evolution}). We calculate the Spearman correlation coefficients between $A_{\rm CAS}$ and the other parameters. As an example, we find that correlation coefficients, between $A_{\rm CAS}$ and $L_{\rm bol}$, reduce from $\sim 0.35$ in bluer bands ($g$, $r$, and $i$) to 0.26 in $z$-band and 0.17 in $y$-band. In addition, for each panel in Figure \ref{fig:host_A_evolution}, the data points for the bluer bands are mostly located above those from the redder bands. As a remark, the $g$-band results \textcolor{black}{are not likely to be greatly affected by residuals from the model point source subtraction} even with a lower host fraction in $g$-band and poorer seeing conditions (see Sec.~\ref{subsec:qualification}). The mean and standard deviation of host fraction for the reliable objects in $grizy$ bands are: $0.26\pm0.22$, $0.37\pm0.22$, $0.45\pm0.21$, $0.49\pm0.20$, and $0.51\pm0.20$, respectively. 

\textcolor{black}{We find a similar trend in Figure \ref{fig:merger_fraction} that the correlations between merger fraction and $L_{\rm bol}$ become weaker in the redder bands. The results in $gri$ bands are roughly consistent, while elevated from the $zy$ bands. The merger fraction of control galaxies are also lower than in $zy$ bands than in $i$-band, which are $14.2\%\pm0.5\%$ and $10.4\%\pm0.4\%$, respectively. These two values are generally higher than the quasar hosts even in the luminous cases. We suggest a possible explanation that, if the quasars are triggered by major mergers, they are more likely gas-rich (``wet") mergers \citep{hopkins2008cosmological}, which are easier to be found in bluer bands than in redder bands.}



\begin{figure*}
\begin{centering}
\includegraphics[width=0.95\textwidth]{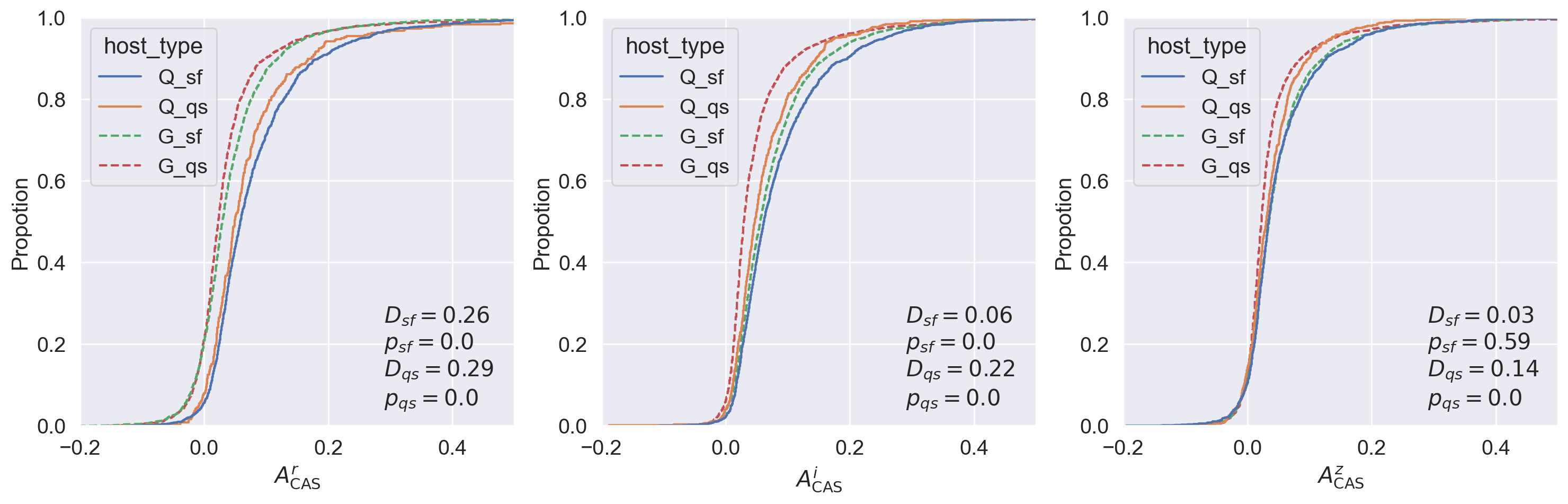}
\caption{\textcolor{black}{Comparison of empirical Cumulative Distribution Functions (eCDF) of $A_{\rm CAS}$ for the $r$ (left panel), $i$ (middle panel), and $z$ (right panel) bands. The sample is separated into four categories (colored curves; see text for details). The Kolmogorov–Smirnov test (K-S test) statistic $D$ and $p$ values are provided in each panel. $D_{\rm sf}$ and $p_{\rm sf}$ are measured between Q\_sf and G\_sf, while $D_{\rm qs}$ and $p_{\rm qs}$ are measured between Q\_qs and G\_qs.} \label{fig:short_A_ecdf}}
\end{centering}
\end{figure*}

In Figure \ref{fig:short_A_ecdf}, \textcolor{black}{we further illustrate the information shown in Figure \ref{fig:QG_A_CAS} by using empirical Cumulative Distribution Functions (eCDF) in $r$, $i$, and $z$ bands.} The curves represent the different samples: blue=star-forming quasar hosts (Q\_sf), orange=quiescent quasar hosts (Q\_qs), green=star-forming galaxies (G\_sf), and red=quiescent galaxies (G\_qs). We show the Kolmogorov–Smirnov test (k-s test) statistic $D$ and $p$ value calculated using \textbf{scipy.stat.ks\_2samp} between quasar hosts and inactive galaxies separated by their types. The k-s $D$ statistic indicates the maximum vertical distance between the two eCDFs. \par

\textcolor{black}{We see a single trend that the $p$ value increases from blue to red and the $D$ value decreases, especially for star-forming hosts. The null hypothesis for $p$ value here is that the two populations are identical in asymmetry, thus larger $p$ value means higher probability that the two populations have the same distribution. Because we keep two digits, $p<0.01$ will be shown as 0. The $D$ statistic indicates the maximum distance between the two eCDFs which can also be visually identified by the solid and dashed curves that are approaching each other from left to right. As in Section~\ref{subsec:QG_A_CAS}, we calculate the excess median quasar host asymmetry over inactive galaxies for the entire sample in $r$, $i$, and $z$ bands, which are $0.030\pm0.002$, $0.017\pm0.002$, $0.006\pm0.002$, respectively. Therefore, we find that the answer to whether quasar hosts are more asymmetric than inactive galaxies has an optical band dependence. The reason for this effect 
will be discussed in detail in Section \ref{subsec:band_interpretation}} \par

\section{Discussion} \label{sec:discussion}

\subsection{The \texorpdfstring{$L_{\rm bol}-A$}\ correlation} \label{subsec:correlations}

\begin{figure}
\begin{centering}
\includegraphics[width=0.45\textwidth]{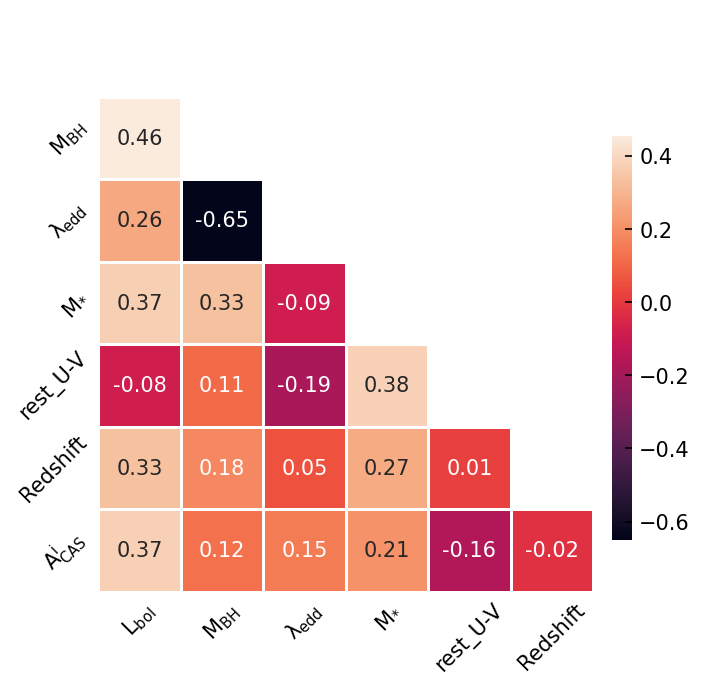}
\caption{Spearman correlation coefficient matrix for seven parameters. We claim that $L_{\rm bol}$ has the strongest correlation with quasar host asymmetry. \label{fig:corr_matrix}}
\end{centering}
\end{figure}

Considering the correlations between asymmetry and quasar properties (Section \ref{subsec:Quasar_A_corr}), we start our discussion on the $i$-band results first and consider the other bands later. We calculate the Spearman correlation coefficient $\rho_s$ between two of the seven parameters. The results are shown as a matrix in Figure \ref{fig:corr_matrix}. Values in each cell indicate $\rho_s$ between the x parameter and y parameter. A lighter color refers to a stronger positive correlation while a darker color indicates a stronger negative correlation. The parameters given in the bottom row are of primary interest which show their dependence of $A_{\rm CAS}^i$. We find that $A_{\rm CAS}^i$ is subtly related to $M_{\rm BH}$ and $\lambda_{\rm edd}$, while more strongly dependent on $L_{\rm bol}$ ($\rho_s=0.37$). \textcolor{black}{Besides the BH properties, the host properties ($M_*$ and rest-frame U-V color) play a lesser role on $A_{\rm CAS}^i$. There is almost no correlation between redshift and $A_{\rm CAS}^i$, as also shown in Figure \ref{fig:host_A_evolution_single}.} For cells in the other rows with $|\rho_s|>0.3$ (Figure \ref{fig:corr_matrix}), the redshift dependence of \textcolor{black}{$L_{\rm bol}$} correlation between $M_*$ and rest-frame U-V color traces the SFR-stellar mass relation \citep{williams2009detection}. The correlations between $M_*$ and BH properties describe the BH-host co-evolution scenario \citep{kormendy2013coevolution}, and agrees with \cite{suh2019multi}, who reported 0.44 and 0.24 Pearson correlation coefficients between $L_{\rm bol}^{\rm X-ray}$ and $M_*$ for AGNs at $z<0.5$ and $0.5<z<1.0$ AGNs. The correlations between the three BH properties are driven by Equations \ref{eq:R_edd} and~ \ref{eq:L_edd}. Considering the mutual dependence between the above-mentioned parameters, we also calculated the partial Spearman correlation coefficient with python package \textsc{pingouin.partial\_corr} between $L_{\rm bol}$ and $A_{\rm CAS}^i$. The effect of $M_{\rm BH}$\footnote{Either $M_{\rm BH}$ or $\lambda_{\rm edd}$ could be controlled, but not both. Otherwise $L_{\rm bol}$ will be fixed}, $M_*$, rest U-V, and redshift are controlled. We find that the partial correlation coefficient remains at 0.31, with a 95\% confidence level between $0.27\sim0.35$. Therefore, we claim that there is a weak independent correlation between quasar $L_{\rm bol}$ and quasar host asymmetry. \par

\begin{figure*}
\begin{centering}
\includegraphics[width=0.95\textwidth]{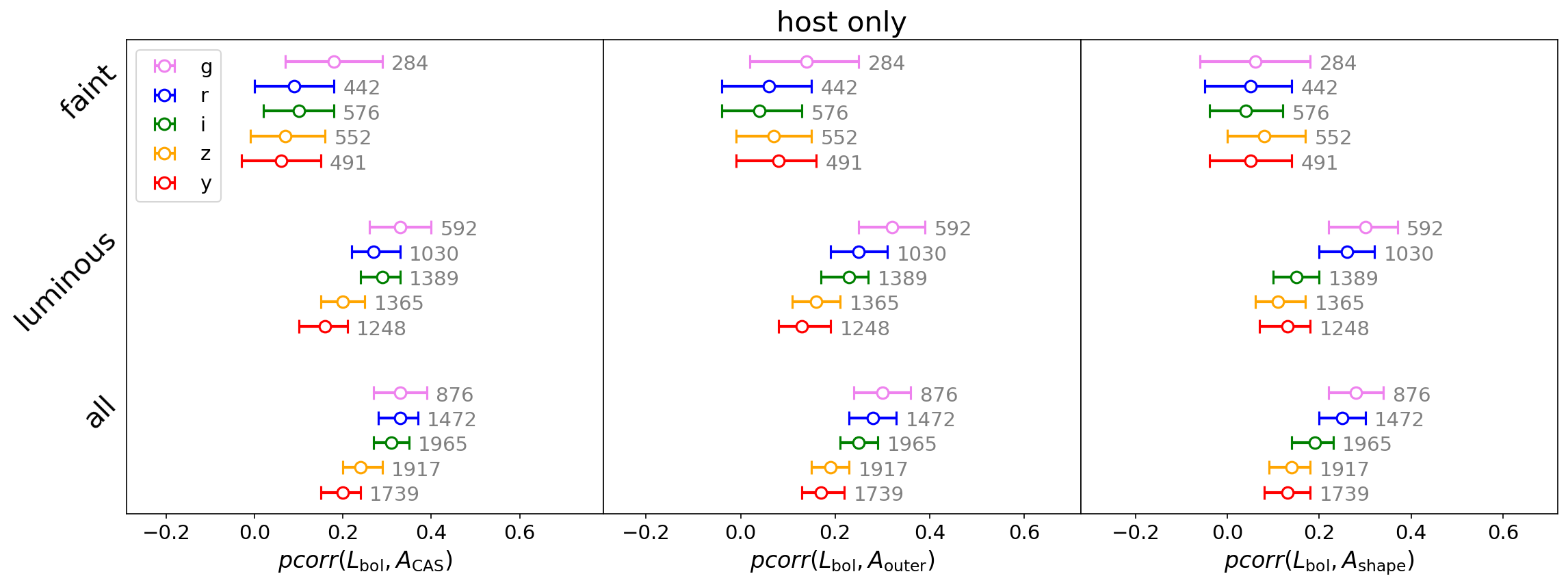}
\end{centering}
\caption{\textcolor{black}{Confidence levels (95\%) on the Spearman partial correlation coefficients between $L_{\rm bol}$ and $A_{\rm CAS}$ (left), $A_{\rm outer}$ (middle), $A_{\rm shape}$ (right). Results for the $grizy$ bands are shown as different colored bars, followed with the numbers of reliable sources in each. Quasars are separated into faint and luminous subcategories with a threshold of $L_{\rm bol} = 10^{45} \mathrm{erg s^{-1}}$. The effects of $M_{\rm BH}$, $M_*$, rest frame U-V color, and redshift are controlled for this partial correlation. 
}
\label{fig:lbol_corr_host}}
\end{figure*}

\textcolor{black}{We then separate the samples into faint and luminous quasars with a threshold of $L_{\rm bol} = 10^{45} \mathrm{erg\ s^{-1}}$ and re-investigate the $L_{\rm bol}-A$ correlation for both subcategories in all five bands using three asymmetry indices. The threshold is decided by the evidence shown in Figure \ref{fig:host_A_evolution_single} (first panel), in which the median asymmetry (green curve) turns up at $\sim45$. We show the 95\% confidence levels of the $L_{\rm bol}-A$ partial correlation Spearman coefficients as colored bars in Figure \ref{fig:lbol_corr_host}. The number of reliable sources in each subcategory is shown in grey. We find that the results for the three asymmetry indices are consistent. The faint subcategory tends to have weaker $L_{\rm bol}-A$ correlation than the luminous sample in all five bands. This indicates that, only when quasars are bright enough, their luminosity could have some correlation with their host. On the other hand, we also see a trend that this correlation is stronger in bluer bands than in redder bands, a reproduction of the results presented in Figure \ref{fig:host_A_evolution}. It could still be explained by the stellar population scenario mentioned above. While this analysis uses "host only" measurements, the results are similar for "host+comp" frames.} \par

Based on these findings, the $L_{\rm bol}-A$ correlation can be interpreted in multiple ways. First, it may mean that high asymmetries cause an increase in $L_{bol}$, consistent with the widely explored merger-trigged AGN accretion scenario. Early simulation studies \citep[e.g.,][]{hernquist1989tidal,barnes1991fueling,di2005energy} showed that galaxy mergers may trigger gas inflow and ignite the central SMBH. This scenario is partly supported by recent observations showing that more luminous AGNs tend to have larger merger fractions \citep{treister2012major,glikman2015major,hong2015correlation,fan2016most,kim2017stellar,alonso2018impact,goulding2018galaxy,ellison2019definitive,gao2020mergers,kim2021hubble}. With correlations between $L_{\rm bol}$, $M_{\rm BH}$, and $\lambda_{edd}$, it is also natural to see an enhanced merger fraction when the BH is more massive \citep{hong2015correlation} or has a higher accretion rate \citep{marian2020significant}. However, given the existing evidence against the correlation between BH properties and merger fraction \citep{kocevski2011candels,villforth2014morphologies,mechtley2016most,villforth2017host,hewlett2017redshift}, and the overall low fraction of mergers among AGNs \citep{gabor2009active,cisternas2010bulk,silverman2011impact}, it is still unclear whether the merger-trigged AGN scenario is enough to explain the correlations described above. As a test, we removed all the mergers identified by $A_{\rm CAS}^i>0.35$ in ``host+comp" frames, and re-calculated $\rho_s$. Interestingly, $\rho_s$ between $A_{\rm CAS}^i$ and $L_{\rm bol}$ remains almost unchanged (0.37). This agrees with Figure \ref{fig:short_A_ecdf} which shows that, for the population with $A_{\rm CAS}$ below 0.35 in $gri$ bands, quasar hosts are also more asymmetric than control galaxies. Therefore, there is something other than major mergers making the quasar hosts more asymmetric than the control galaxies, and it is correlated to $L_{\rm bol}$. It is still possible to be merger-related (i.e., minor or late-stage mergers) because the $A>0.35$ criteria mainly selects major mergers. Recent semi-analytic models such as $v^2 G C$ \citep{shirakata2019new} showed that minor mergers could trigger $30\% \sim 90\%$ of AGNs at $z<1$ from the brightest population to the faintest (also see \cite{fanidakis2012evolution,menci2014triggering,griffin2019evolution}). Future work with Gini and $M_{20}$ measurements \citep{lotz2004new} on the same sample set will help us to test this scenario.\par

Alternatively, it can be $L_{\rm bol}$ that causes increased asymmetries ($A_{\rm CAS}^i$). It is known that morphological parameters trace the star formation activity to some level (\cite{conselice2000asymmetry, conselice2003relationship}, also discussed). The star-forming main sequence (SFMS, SFR-stellar mass relation) has been reported to have a scatter of $\sim 0.3$ dex \citep{daddi2007multiwavelength,speagle2014highly}. This means besides the stellar mass, some other properties of the galaxies are related to the SFR. Using $\sim 3700$ SDSS Stripe 82 star-forming galaxies (SFGs), \cite{yesuf2021important} applied the mutual information framework to study the relevance of several structural parameters on the offset of specific SFR from SFMS. They found that asymmetry provides the most information on the offset residual, thus is the second important parameter (next to $M_*$) to the SFR of galaxies. This is supported by the negative correlation between the color and asymmetry that we see in Figure \ref{fig:corr_matrix}, indicating bluer quasar hosts are more asymmetric. Therefore, here we roughly translate the $L_{\rm bol}-A$symmetry correlation (after removing the effect of $M_*$) to the positive AGN feedback scenario, in which the BH activity may trigger SFR \citep{ishibashi2012active,silk2013unleashing}. This scenario may also explain our findings in Section \ref{subsec:band_interpretation}. The correlation is stronger in bluer bands because it is the younger stellar population that directly relates to the AGN activity, while the older population is less affected. And if the AGN is more powerful (luminous), it will have stronger feedback to form stars. Especially in quiescent quasar hosts, the newly-born stars are separated from the old population, while in star-forming quasar hosts, they could be mixed up with pre-quasar phase star-forming activities \textcolor{black}{(Figures \ref{fig:QG_A_CAS} and~\ref{fig:short_A_ecdf})}. Also, if the quasar host is less massive, it would be easier for the quasar to affect it \textcolor{black}{(Figure \ref{fig:heat_host})}. Studies have directly tested this connection using different SFR estimators including far IR luminosity, SED fitting, and [O\,{\sc ii}]-[O\,{\sc iii}] emission lines \citep{hickox2014black,lanzuisi2017active,zhuang2020interplay}. However, our HSC data does not cover far IR and five-band optical data is not reliable enough to fit a good SED model for SFR estimation. The \cite{rakshit2020spectral} catalog does not include the [O\,{\sc ii}] emission line. We leave the estimation of SFR of our quasar hosts for a future work.\par

Lastly, the correlation between $L_{\rm bol}$ and $A_{\rm CAS}^i$ (considering it here as an indicator of SFR) could be driven by a third parameter that is mutually related to both and not considered in Figure \ref{fig:corr_matrix}. One possibility is the molecular gas mass $M_{\mathrm{H}_{2}}$. It is suggested that the growth of BH and star formation activity could be fueled by the same cold gas reservoir \citep{springel2005simulations,hopkins2006fueling}. For example, \cite{shangguan2020agn} studied CO lines of 40 Palomar-Green (PG) quasars at $z<0.3$ observed by ALMA. They estimated the molecular gas mass of the quasar hosts according to the CO-to-$\mathrm{H_2}$ conversion factor ($\rm \alpha_{CO}$). They found the correlation between IR luminosity $L_{\rm IR}$, which is used as an indicator of SFR, and AGN luminosity $\lambda L_{\lambda}(5100 \text{\AA})$ disappears after removing the dependence of $L_{\rm IR}$ on $L_{\rm CO}$, an indicator of $M_{\mathrm{H}_{2}}$. Latter work done by \cite{zhuang2021black} investigated 453 SDSS DR7 quasars at $0.3<z<0.35$. They estimated the molecular gas of quasar hosts using a new method developed by \cite{yesuf2019dirt} based on dust extinction and gas-phase metallicity. They also found that the correlation between BH activity (they used BHAR) and SFR have mutual dependence on $M_{\mathrm{H}_{2}}$. And even after removing the effect of $M_{\mathrm{H}_{2}}$, a weak correlation still exists. \cite{yesuf2020gas} also applied the same method and found that the molecular gas fraction and the Eddington ratio have moderately strong correlations in AGN-dominated galaxies. However, we cannot test this gas-driven scenario with either method. We do not have CO line information about our quasars, and the \cite{yesuf2019dirt} method is also not applicable for the whole sample. Because the dust extinction was traced by $\mathrm{H\alpha/H\beta}$, while not all of our quasar spectra cover H$\alpha$. Yesuf et al. (in preparation) is performing a test on this method using $\sim 2000$ HSC Type-1 AGNs at $z<0.35$. \par



In summary, there are at least three possible scenarios that can explain the $L_{\rm bol}-A$symmetry correlation. Nevertheless, it is difficult to distinguish between them with our current information in hand. More detailed classification of mergers, and more reliable measurements of SFR and gas mass are required in future work.

\subsection{Merger rates}

\begin{figure*}
\begin{centering}
\includegraphics[width=15cm]{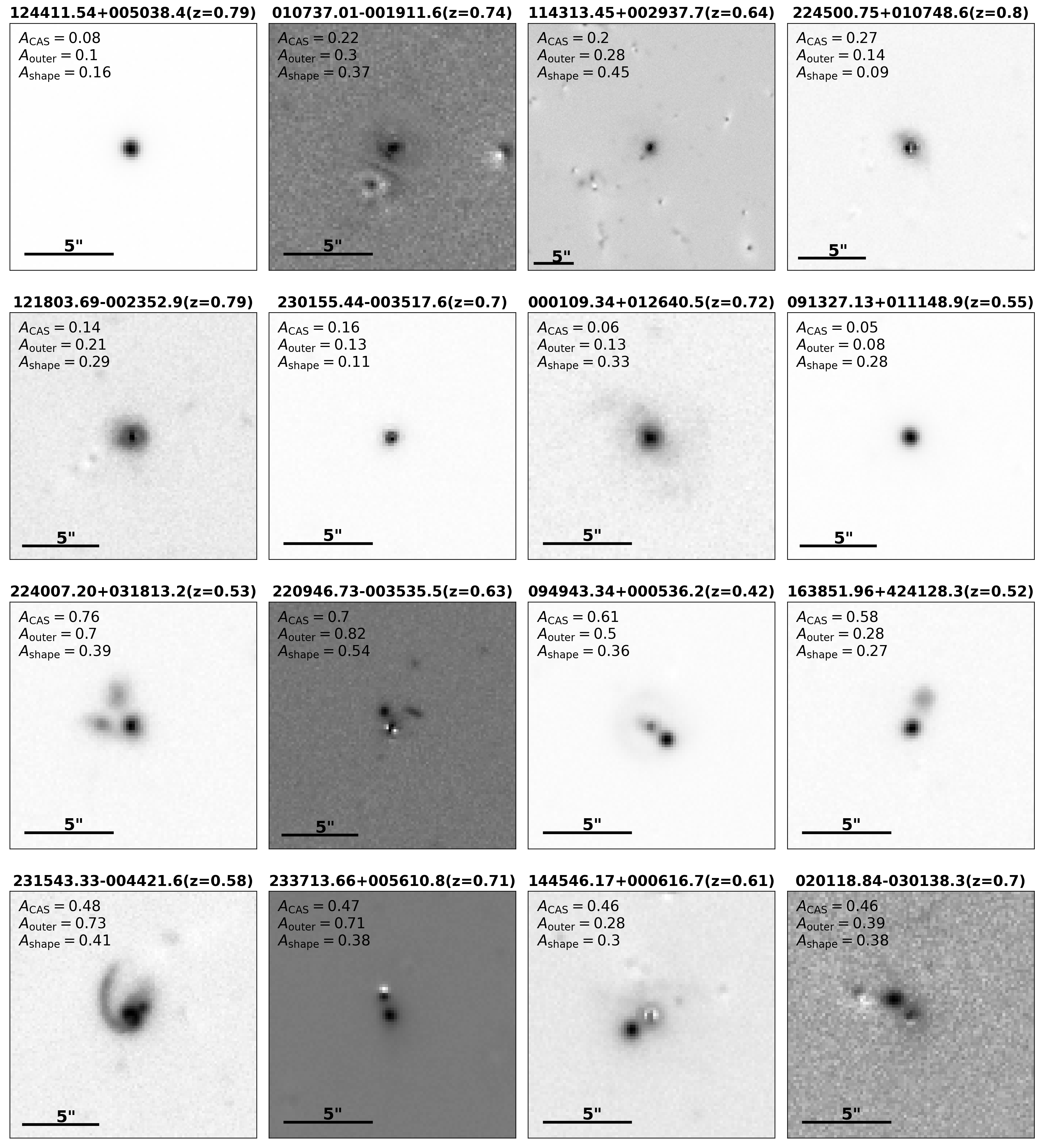}
\caption{$i$-band images after removing point source and companions for 8 most luminous cases (upper two rows) and 8 most asymmetric cases (bottom two rows). Each source is titled with its SDSS name and spectroscopic redshift. $A_{\rm CAS}$, $A_{\rm outer}$, and $A_{\rm shape}$ values are shown for each source. We find that the most luminous quasars and most asymmetric quasar hosts are two different populations. The asymmetry values of the former are mainly contributed by intrinsic processes and typically around 0.2, while that of the later are mainly contributed by extrinsic processes and typically above 0.3.
\label{fig:montages_luminous_asymmetry}}
\end{centering}
\end{figure*}

\textcolor{black}{Considering the best-fit relations in Figure~\ref{fig:merger_fraction}, the intersection between the merger fractions of quasars and control galaxies is located at $L_{\rm bol}\sim45$, similar to the turning point we reported in Figure \ref{fig:host_A_evolution_single}. Below this luminosity, the merger fractions of quasars are lower than that of control galaxies. The merger fraction of the entire control galaxy sample in $i$-band is $16.8\pm0.5\%$ (see Sec.~\ref{subsec:QG_A_CAS} for asymmetry values of control galaxies), which is shown as the shaded grey line in Figure \ref{fig:merger_fraction}. We suggest a possible explanation that low luminosity quasars are preferentially formed by minor mergers \citep{shirakata2019new}, or secular processes such as the influence of bars \citep{crenshaw2003host, ohta2007bar} and disk instabilities \citep{hirschmann2012origin}, or found as fading quasars in merger remnants \citep{bennert2008evidence}. Therefore, low luminosity quasars are a biased population that is less likely to be classified as mergers using the asymmetry values, which is only sensitive to major mergers \citep{conselice2014evolution}. 
On the other hand, luminous quasars ($L_{\rm bol}>45$) have an excess of mergers up to $\sim15\%$ compared to the control galaxies. Referring to examples shown in Figure \ref{fig:montages_luminous_asymmetry}, seven of the eight most asymmetric cases have $L_{\rm bol}>45$. Our findings that luminous quasars are more likely to be in mergers reflects the high $L_{\rm bol}$ values for those within the high asymmetry tail of the distribution as indicated by the orange data points in Figure \ref{fig:host_A_evolution_single}. Note that the most luminous bin in Figure \ref{fig:merger_fraction} is roughly the second most luminous bin in Figure \ref{fig:host_A_evolution_single}, we did not show the merger fraction for the most luminous quasars with $L_{\rm bol}>46$ here due to the poor statistics (error bars covering $\sim20\%$). Therefore, the claim here that luminous quasars are more likely to be mergers does not conflict with the claim in Section \ref{subsec:Quasar_A_corr} that most of the most luminous quasars are not mergers. It's worth noting that our best-fit relation (Fig. \ref{fig:merger_fraction}) reaches 0.35 at most thus never approaches unity.} 

\subsection{\textcolor{black}{Interpretation of optical band dependencies and its impact on determining merger fraction}} \label{subsec:band_interpretation}
\begin{figure*}
\begin{centering}
\includegraphics[width=0.85\textwidth]{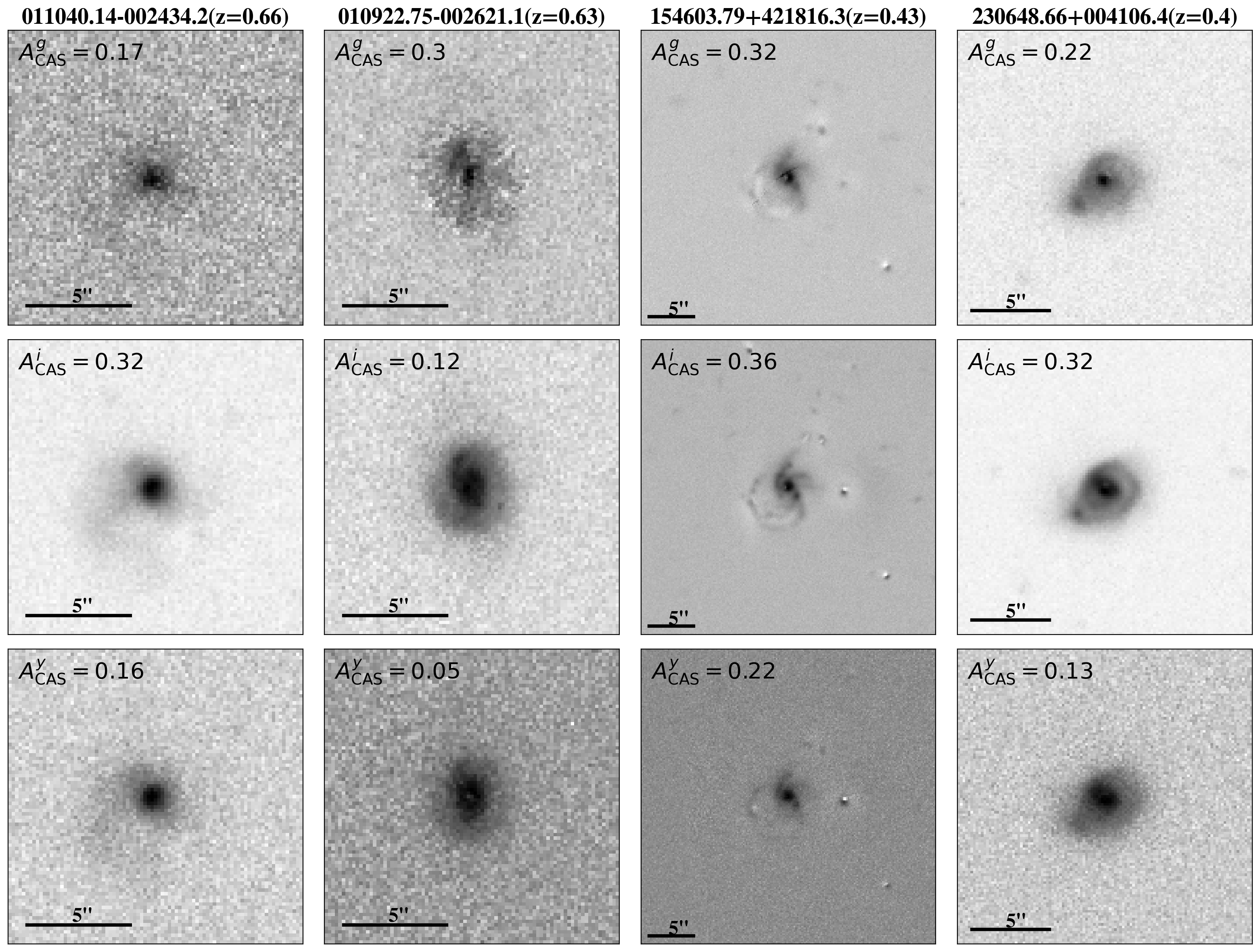}
\end{centering}
\caption{Examples of four ``host only" Type-1 quasar hosts (SDSS name and redshift are titled on top of each column), whose CAS asymmetries were measured in \textcolor{black}{$giy$ bands from top to bottom}. We show that the variation of CAS asymmetry values over bands could be caused by band-dependent, faint tidal tails, bad quasar/AGN subtraction, the band-dependent prevalence of young stellar populations, and blurred secondary galaxies. \label{fig:montages_A_discrepancy}}
\end{figure*}
Previous works have not reached a clear consensus on whether the merger rate of quasar hosts is higher than inactive galaxies over a broad range of redshift and across the demographics of the AGN population. For example, \citet{marian2020significant} finds an enhancement of the merger fraction of AGNs at the highest Eddington ratios at $z<2$. As well, \citet{goulding2018galaxy} finds a clear excess of mergers associated with luminous obscured AGNs in the HSC-SSP survey. However, \cite{villforth2017host} reports that even the most luminous AGNs do not have an enhanced rate of disturbance compared to inactive galaxies. Therefore, the tension is not yet fully resolved. \textcolor{black}{Beyond the luminosity dependence that we have shown in Figure \ref{fig:merger_fraction}, we suggest that the band dependence could be another factor to reconcile the differences seen in the literature. }\par
To demonstrate the dependence on a given optical band, we show four cases of ``host only" quasar images whose $A_{\rm CAS}$ values differ in blue and red bands (Figure \ref{fig:montages_A_discrepancy}). For SDSS J011040.14-002434.2, the excess of $A_{\rm CAS}^i$ is mainly driven by the better image quality in this band (Section \ref{subsec:qualification}). In the second case, SDSS J010922.75-002621.1, we find the excess in $gr$ band could be artificial due to the subtraction residual in the center. In a third case, SDSS J154603.79+421816.3, the excess in bluer band is caused by better-resolved spiral arms and star-forming regions. In a fourth case, SDSS J230648.66+004106.4 is probably a minor merger. The secondary galaxy is evident in bluer bands, while blurred with the main galaxy in $y$-band. \par
Correspondingly, we consider four general reasons for the excess asymmetry in the bluer bands. First, it could be a technical bias related to the image quality in different bands. Specifically, the S/N and resolution are two main factors that may affect the asymmetry measurements. In Appendix \ref{sec:refinements}, we show how we correct the S/N issue according to a simulation work by \cite{thorp2021towards}. While not correcting for resolution-based issues, we see in Figure \ref{fig:Resolution_A} that the excess of asymmetry in bluer bands still exists even in smaller spatial bins. The values in Figure \ref{fig:montages_A_discrepancy} have already been corrected for S/N. Second, it could be a systematic effect caused by our 2D image decomposition method using \textsc{GaLight}. However, the excess still exists as seen from a check of the results based on the ``host+comp" frames which include the companions. Considering the subtraction of point sources, because the residual mainly affects the central region, $A_{\rm outer}$ can mitigate this effect to some level. And since $A_{\rm shape}$ flattens all the values, it is less sensitive to the poor residuals. Therefore, we check $A_{\rm outer}$ and $A_{\rm shape}$ and find they also have excesses in the bluer bands. Third, it could be caused by the stellar population. The bluer bands capture light from a younger stellar population, while the redder bands are more sensitive to the older stars. In that sense, the excess in bluer bands would indicate that the younger stellar population is more unevenly distributed as compared to the older stellar population. This has already been pointed out by early works \citep{conselice1997symmetry, conselice2000asymmetry}, in which they refer to it as "flocculent" asymmetry. Fourth, it could be a systematic bias in identifying mergers in different bands. For example, when the secondary galaxy or tidal features are dominated by young stars that are newly triggered by the merger, the system will more likely be identified as a merger. This is somewhat entangled with the "flocculent" asymmetry caused by star-formation activity but mainly driven by the dynamics. Spectroscopic selection of mergers such as being applied in \cite{ellison2013galaxy,ellison2019definitive} will yield purer merger samples to deal with the systematic effects.
\textcolor{black}{Considering that most of our samples are non-mergers, and they usually do not have significant issues with subtraction residuals (otherwise rejected), we suggest that "flocculent" asymmetry to be the main cause of the statistical excess of asymmetry in the bluer bands.} \par

We then compared our results with previous works that have studied the merger rate of Type-1 AGNs at various luminosities (Table \ref{tab:works}). None of these works discussed the "flocculent" asymmetry because most of them only used one photometric band. Here, we can explore whether the asymmetry is affected by the stellar population by comparing results at a common rest-frame wavelength (Column 6). All of these works compared the merger fractions between AGNs and control inactive galaxies, which are listed in Columns 7 and 8. Because the definition of mergers vary in different studies, instead of the fraction itself, comparison between $f_{\rm merger}^Q$ and $f_{\rm merger}^G$ are more meaningful. In that sense, we find these works follow such a pattern: when considering stellar populations at $\lambda_{\rm eff}^{\rm rest} < 5200 \text{\AA}$ (our HSC study in $gri$ bands, \cite{gabor2009active,cisternas2010bulk,koss2010merging,boehm2013agn,mechtley2016most,marian2019major,marian2020significant}), they tend to find more or less an excess of $f_{\rm merger}^Q$ over $f_{\rm merger}^G$. When sensitive to $5200 \text{\AA} < \lambda_{\rm eff}^{\rm rest} < 9000 \text{\AA}$ (This work in $zy$ band, \cite{cotini2013merger,ellison2019definitive,zhao2022relation}), the results are mixed. At $\lambda_{\rm eff}^{\rm rest} > 9000 \text{\AA}$ \citep{villforth2014morphologies,villforth2017host}, so far no excess of $f_{\rm merger}^Q$ is reported. Rest-frame wavelengths around 5200 \text{\AA} are roughly the turning point in the spectra of G and K stars \citep{yan2019sdss}. Its blue ward emission is contributed by younger O, B, A, and F stars. Up to 9000 \text{\AA}, it is contributed by all types of stars. While beyond 9000 \text{\AA}, the M stars dominate the spectra. \textcolor{black}{We reemphasize our point mentioned in Section \ref{subsec:band} and Figure \ref{fig:merger_fraction} that, if a quasar is formed in major merger, such a merger is more likely to be gas-rich \citep{hopkins2008cosmological}, thus have young stellar population. Such a merger is easier to be classified either with quantification method or visual inspection in bluer bands than in redder bands.} We do not deny a possible intrinsic redshift evolution of the asymmetry at higher redshift, since we only showed a flattened curve at $z<0.8$ (Figure \ref{fig:host_A_evolution}). Also, we clarify that our results are only for Type-1 quasars.



\begin{landscape}
 \begin{table}
 \caption{\textcolor{black}{Merger fractions of quasar hosts and normal galaxies}}
  \label{tab:works}
  \begin{tabular}{cccccccccc}
    \hline
    works & $N_{\rm AGN}$ & observed band & $z_{\rm range}$ & $<z>$ & $\lambda_{\rm eff}^{\rm rest} (\text{\AA})$ & \textcolor{black}{log$L_{\rm bol}$} & $f_{\rm merger}^Q$ & $f_{\rm merger}^G $ & classifier\\
    (1) & (2) & (3) & (4) & (5) & (6) & (7) & (8) & (9) & (10) \\
    \hline
    This work & 886 & HSC\_g & $0.2-0.8$ & 0.49 & 3194 & $43.1-46.4$ & $15.8\%\pm1.4\%$ & \textcolor{black}{$14.4\%\pm0.7\%$} & Asymmetry\\
    This work & 1488 & HSC\_r & $0.2-0.8$ & 0.54 & 3984 & $43.1-46.4$ & $16.0\%\pm1.0\%$ & \textcolor{black}{$14.7\%\pm0.6\%$} & Asymmetry\\
    This work & 1987 & HSC\_i & $0.2-0.8$ & 0.60 & 4801 & $43.1-46.4$ & $16.2\%\pm0.9\%$ & \textcolor{black}{$16.8\%\pm0.5\%$} & Asymmetry\\
    This work & 1939 & HSC\_z & $0.2-0.8$ & 0.60 & 5567 & $43.1-46.4$ & $10.3\%\pm0.7\%$ & \textcolor{black}{$14.2\%\pm0.5\%$} & Asymmetry\\
    This work & 1759 & HSC\_y & $0.2-0.8$ & 0.58 & 6177 & $43.1-46.4$ & $8.0\%\pm0.7\%$ & \textcolor{black}{$10.4\%\pm0.4\%$} & Asymmetry\\
    \hline
    \cite{gabor2009active} & 19 & ACS\_F814W & $0.3-1.0$ & $\sim 0.7$ & 4690 & $43.3-45.3$ & 68\% $^a$ & $\sim 25\%$ $^a$ & Asymmetry \\
    \cite{cisternas2010bulk} & 83 & ACS\_F814W & $0.3-1.0$ & 0.8 & 4429 & $43.6-46.3$ & $15.0\%\pm8.8\%$ & $12.6\%\pm6.5\%$ & Visual \\
    \cite{koss2010merging} & 72 & SDSS\_gri & $< 0.05$ & 0.03 & 4536 $^b$ & $\sim44$ & 18\% & 1\% & Visual \\
    \cite{kocevski2011candels} & $72^*$ & WFC3\_F160W $^c$ & $1.5-2.5$ & $\sim 2.0$ & 5093 & $43.3-45.3$ & $16.7_{-3.5}^{+5.3}\%$ & $15.5_{-2.2}^{+2.8}\%$ & Visual \\
    \cite{boehm2013agn} & 21 & ACS\_F606W & $0.5-1.1$ & 0.71 & 3397 & $43.3-45.3$ & $0\%-65.5\%$ $^d$ & $20.2\%-52.6\%$ & Both \\
    \cite{cotini2013merger} & $59^*$ & SDSS\_r & $0.003-0.03$ & $\sim 0.02$ & 6021 & $\sim44$ & $20_{-5}^{+7}\%$ & $4_{-1.2}^{+1.7}\%$ & Both \\
    \cite{villforth2014morphologies} & $60^*$ & WFC3\_F160W & $0.5 -0.8$ & $\sim 0.65$ & 9259 & $42.3-45.8$ & $<6\%$ & $\sim 10\%$ & Both \\
    \cite{mechtley2016most} & 19 & WFC3\_F160W & $1.9-2.1$ & 2.0 & 5093 & $46.7-47.7$ & $39\%\pm11\%$ & $30\%\pm5\%$ & Visual \\
    \cite{villforth2017host} & $20^*$ & WFC3\_F160W & $0.5-0.7$ & 0.62 & 9431 & $45.4-47.2$ & $\sim 25\%$ & $\sim 25\%$ & Both \\
    \cite{ellison2019definitive} & 1124 & CFIS\_r & $0- 0.25$ & $\sim 0.1$ & 5760 & $41.8-46.3$ & $7.7\% -30.9\%$ $^e$ & $2.9\%-16.0\%$ & Visual \\
    \cite{marian2019major} & 21 & WFC3\_F160W & $1.8-2.2$ & 1.95 & 5179 & $46.5-47.2$ & $24\%\pm9\%$ & $19\%\pm4\%$ & Visual\\ 
    \cite{marian2020significant} & 17 & FORS2\_B & $0.09-0.19$ & 0.15 & 3687 & $45.3-46.9$ & $41\%\pm12\%$ $^f$ & $8\%\pm6\%$ & Visual\\
    \cite{marian2020significant} & 17 & FORS2\_V & $0.09-0.19$ & 0.15 & 4723 & $45.3-46.9$ & $41\%\pm12\%$ & $8\%\pm5\%$ & Visual\\
    \cite{zhao2022relation} & 245 & PS1\_i & $0.04-0.15$ & 0.09 & 6883 & $43.4-43.8$ & 3.7\% & 3.2\% & Asymmetry\\
    \hline
  \end{tabular}\par
    \bigskip List of works that have been working on the comparison between merger fractions of Type-1 AGN host and inactive galaxies.\\
    \textcolor{black}{Column (2): Number of Type-1 AGNs used in the work. If the Type-1 and Type-2 were not separated (usually selected from X-ray), we count the total number and mark it with a *.}\\
    Column (3): The bands from which the imaging was taken and classified.
    Column (4): Redshift range of the AGNs.\\
    Column (5): The median redshift of the samples. If it is not provided in the work, we roughly estimate it from the distribution plots.\\
    Column (6): The effective wavelength midpoint of the band divided by $1+z_{\rm median}$. Information on the bands is taken from the SVO filter profile service.\\
    \textcolor{black}{Column (7): Range of bolometric luminosities of the AGNs. When only X-ray luminosity is provided in the work, we convert it to $L_{\rm bol}$ with a typical correction factor $L_{\rm bol}/L_{\rm X (2-10 keV)}=20$ according to \cite{lusso2012bolometric} or $L_{\rm bol}/L_{\rm X (14-195 keV)}=8$ according to \cite{koss2017bat}. Some works also use $L_{\rm [O III]}$, we convert it with a correction factor $L_{\rm bol}/L_{\rm [O III]}=600$ according to \cite{kauffmann2009feast}.}\\
    Column (8-9): Merger fraction of AGNs (or Quasars) and matched inactive galaxies. For this work, the uncertainties are the Poisson errors, i.e., $\sqrt{N_{\rm merger}}/N_{\rm total}$.\\
    Column (10): Methods being used to classify mergers. "Both" means both visual and quantitative methods were used.\\
    a: We only count for their Type-1 AGNs (class "X1") with good fittings. The mergers are classified as $A>0.35$, although not used in their work. They did not provide the precise merger fraction of the control galaxies, or the data. So we roughly counted $f_{\rm merger}^G $ from their Figure 10 X-ray panel. However, this comparison might not be fair since the Type-1 AGNs are a subset of their X-ray AGNs, while the control galaxies are matched to all X-ray AGNs.\\
    b: They used composite gri images for classification, we see from Figure 1 that the interacting features are mostly blue. Thus, we note the $\lambda_{\rm eff}^{\rm rest}$ for SDSS\_g band here.
    c: They also had F125W band in their data, but was not used in the results.\\
    d: According to their Table 1 and 2. The lower limit assumes only their major mergers are real, while the upper limit adds all their unclear cases as well. We take the average values of their 4 criteria as the number of mergers. \\
    e: According to their Table 1, the lower limit counts the interacting pairs only, while the upper limit includes the post-mergers. \\
    f: According to their Figure 3, cut-off rank = 10. The same for the V band below.
 \end{table}
\end{landscape}

\section{Conclusions}

We have measured asymmetry indices of 2424 quasar host galaxies at $0.2<z<0.8$ from the SDSS DR14 catalog using all five broad-band optical images from the HSC SSP. This represents the largest sample of Type-1 quasar hosts with asymmetry measurements. We determine the dependence of asymmetry on intrinsic quasar and host properties including systematic differences between optical bands. Control galaxies are selected by matching the redshift and stellar mass of the quasar hosts. The \textsc{GaLight} image decomposition tool is used to subtract the point source from the quasar images. We use two types of quasar host images, either with or without the nearby companions removed based on model fitting. Both are then analyzed using \textsc{statmorph} to measure the $A_{\rm CAS}$, $A_{\rm outer}$, and $A_{\rm shape}$ values. \textcolor{black}{Simulations with mock quasars do not show a bias in the asymmetry measurements caused by the quasar subtraction. After applying the \textsc{statmorph} flags, we keep $\sim 80\%$ of the total sample in $i$-band ($37\%-80\%$ in other bands) of which only $\sim 2\%$ of these sources are affected by significant subtraction issues (Section \ref{subsec:qualification}). Based on these sources, we find the following:}

\begin{itemize}

\item \textcolor{black}{The asymmetry indices of quasar hosts are correlated with $L_{\rm bol}$ primarily due to a broad tail of higher asymmetries at higher quasar luminosities (Sec.~\ref{subsec:Quasar_A_corr} and ~\ref{subsec:merger_frac}).  There is a clear upturn in the median asymmetries binned in luminosity, hence merger rates, at $L_{\rm Bol}=45$. This correlation is stronger in the bluer bands.} 


\item Quasar hosts are more asymmetric when they harbor more massive and more active black holes (i.e., higher Eddington rate), and when their host galaxies are more massive and bluer (Section \ref{subsec:Quasar_A_corr}).

\item Type-1 quasar hosts are \emph{slightly} more asymmetric than inactive galaxies with a median difference of $\sim$ 0.02 in $i$-band with a significance of 9.4$\sigma$. The overall merger fractions of quasar hosts and inactive galaxies are both around $\sim8.0-17.1\%$.

\item An optical band (wavelength) dependence is an essential factor in (1) the excess of asymmetry in quasar hosts over control galaxies, (2) the correlations between the physical properties, and (3) merger ratios. Bluer bands ($gri$) generally have larger asymmetry values than the redder bands ($yz$). Such an effect is stronger for quasar hosts than the inactive galaxies. In bluer bands, the differences between quasar hosts and inactive galaxies is larger, and the correlations between quasar properties and asymmetry also become stronger. While all the correlations are weaker in redder bands (Section \ref{subsec:band}).

\item Bringing our results to common rest-frame wavelengths and comparing to previous works, we find that the discrepancies on whether quasar hosts have merger fractions higher than inactive galaxies could be explained partly by the stellar populations of the quasar host galaxies. At rest-frame wavelengths bluer than 5200 \text{\AA}, $f_{\rm merger}^Q$ is more or less larger than $f_{\rm merger}^G$. When the rest wavelengths are between 5200 \text{\AA} and 9000 \text{\AA}, the results are mixed. Beyond 9000 \text{\AA}, no excess of $f_{\rm merger}^Q$ is reported (Section \ref{subsec:band_interpretation}).

\end{itemize}

\section*{Acknowledgements}
The Hyper Suprime-Cam (HSC) collaboration includes the astronomical communities of Japan and Taiwan, and Princeton University.  The HSC instrumentation and software were developed by the National Astronomical Observatory of Japan (NAOJ), the Kavli Institute for the Physics and Mathematics of the Universe (Kavli IPMU), the University of Tokyo, the High Energy Accelerator Research Organization (KEK), the Academia Sinica Institute for Astronomy and Astrophysics in Taiwan (ASIAA), and Princeton University.  Funding was contributed by the FIRST program from the Japanese Cabinet Office, the Ministry of Education, Culture, Sports, Science and Technology (MEXT), the Japan Society for the Promotion of Science (JSPS), Japan Science and Technology Agency (JST), the Toray Science  Foundation, NAOJ, Kavli IPMU, KEK, ASIAA, and Princeton University.
\par
This paper is based [in part] on data collected at the Subaru Telescope and retrieved from the HSC data archive system, which is operated by Subaru Telescope and Astronomy Data Center (ADC) at NAOJ. Data analysis was in part carried out with the cooperation of Center for Computational Astrophysics (CfCA) at NAOJ.  We are honored and grateful for the opportunity of observing the Universe from Maunakea, which has the cultural, historical and natural significance in Hawaii.
\par
This paper makes use of software developed for Vera C. Rubin Observatory. We thank the Rubin Observatory for making their code available as free software at \text{http://pipelines.lsst.io/}. 
\par
The Pan-STARRS1 Surveys (PS1) and the PS1 public science archive have been made possible through contributions by the Institute for Astronomy, the University of Hawaii, the Pan-STARRS Project Office, the Max Planck Society and its participating institutes, the Max Planck Institute for Astronomy, Heidelberg, and the Max Planck Institute for Extraterrestrial Physics, Garching, The Johns Hopkins University, Durham University, the University of Edinburgh, the Queen’s University Belfast, the Harvard-Smithsonian Center for Astrophysics, the Las Cumbres Observatory Global Telescope Network Incorporated, the National Central University of Taiwan, the Space Telescope Science Institute, the National Aeronautics and Space Administration under grant No. NNX08AR22G issued through the Planetary Science Division of the NASA Science Mission Directorate, the National Science Foundation grant No. AST-1238877, the University of Maryland, Eotvos Lorand University (ELTE), the Los Alamos National Laboratory, and the Gordon and Betty Moore Foundation.
\par
We thank the anonymous reviewers for their insightful comments on an earlier draft of this work. S.T. thanks Mallory D. Thorp for her advice on the application of corrections to asymmetry values. J.D.S. is supported by the JSPS KAKENHI Grant Number JP22H01262, and the World Premier International Research Center Initiative (WPI Initiative), MEXT, Japan. H. Yesuf was supported by JSPS KAKENHI Grant Number JP22K14072 and the Research Fund for International Young Scientists of NSFC (11950410492). X.D. is supported by JSPS KAKENHI Grant Number JP22K14071. C.B. gratefully acknowledges generous support from the Natural Sciences and Engineering Research Council of Canada through their post-doctoral fellowship program. \par
\textcolor{black}{Software: numpy \citep{harris2020array}; matplotlib \citep{Hunter:2007}; pandas \citep{reback2020pandas}; seaborn \citep{Waskom2021}; scikit-learn \citep{scikit-learn}; TOPCAT \citep{taylor2011topcat}; \textsc{GaLight} \citep{ding2021galaxy}; \textsc{photutils} \citep{larry_bradley_2022_6825092}; \textsc{statmorph} \citep{rodriguez2019optical}.}\par
Database: SVO filter profile service, \text{http://svo2.cab.inta-csic.es/theory/fps/}

\section*{Data Availability}
\begin{table*}
\caption{Asymmetry measurements for quasar hosts, complete version is available online \label{tab:A_quasar}}
\begin{tabular}{cccccccc}
\hline
name & host\_A\_i & host\_oA\_i & host\_sA\_i & host\_s2n\_i & host\_res\_i & host\_flag\_i & ...\\
\hline
000002.50+021818.9 & 0.129 & 0.334 & 0.233 & 9.187 & 1.56 & 0 & ...\\
000010.97+005653.3 & 0.031 & 0.038 & 0.069 & 16.818 & 0.77 & 0 & ... \\ 
000017.74+020027.9 & 0.065 & 0.143 & 0.445 & 29.709 & 1.67 & 0 & ... \\ 
000017.88+002612.6 & 0.074 & 0.117 & 0.132 & 14.253 & 1.42 & 0 & ... \\ 
000022.05+011742.5 & 0.047 & 0.275 & 0.567 & 61.347 & 1.11 & 0 & ... \\ 
... & ... & ... & ... & ... & ... & ... & ... \\
\end{tabular} \par
\end{table*}

\begin{table*}
\caption{Asymmetry measurements for control galaxies, complete version is available online \label{tab:A_galaxy}}
\begin{tabular}{cccccccc}
\hline
name & host\_A\_i & host\_oA\_i & host\_sA\_i & host\_s2n\_i & host\_res\_i & host\_flag\_i & ...\\
\hline
43128210455553649 & 0.091 & 0.151 & 0.268 & 5.559 & 1.14 & 1 & ...\\  
42072271270998533 & 0.036 & 0.025 & 0.146 & 5.29 & 1.21 & 1 & ...\\
44191880581244422 & 0.038 & 0.086 & 0.236 & 7.529 & 1.04 & 1 & ...\\
40972617909296239 & 0.029 & 0.042 & 0.102 & 10.603 & 0.94 & 0 & ...\\
40985945192811690 & 0.084 & 0.127 & 0.153 & 4.13 & 1.4 & 0 & ...\\
... & ... & ... & ... & ... & ... & ... & ... \\
\end{tabular} \par
\end{table*}
The measurements of asymmetries for each source, including the quasar hosts and control inactive galaxies are available online. The entire catalog included the physical parameters of the quasars, we refer the readers to Table 1 of \cite{li2021sizes} for the meaning of those lines. The asymmetry measurement results are appended from the 36th column to the 100th column. Basically, the columns are named as ``\{sersic/host\}\_\{A/oA/sA/s2n/res/flag\}\_\{g/r/i/z/y\}", where ``sersic" stands for the measurement for ``host+comp" frames, ``host" stands for that of ``host only" frames. ``A", ``oA", and ``sA" represent the CAS asymmetry, outer asymmetry, and shape asymmetry, respectively. ``s2n" stands for the signal-to-noise ratio per pixel of the host galaxies measured by \textsc{statmorph}, ``res" stands for the resolution of the galaxy estimated with half-light radius and FWHM. These two parameters are used in Appendix \ref{sec:refinements} to qualify the corrections of asymmetry values using simulation. All the asymmetry values in our data table have applied the corrections. Finally, ``flag" stands for the \textsc{statmorph} flag, 0=good and 1=bad (Section \ref{subsec:qualification}). The empty cells in the data table are the failures at the stage of deblending and masking (Section \ref{subsec: MD}). These make up $2\times6\times5 = 60$ columns, a short example for ``host only" frames in $i$ band is shown in Table \ref{tab:A_quasar}. We also added five columns for the sky background asymmetry measured for each source, which is named ``sky\_\{g/r/i/z/y\}". These are the $A_{\rm bkg}/1.2$ values mentioned in Section \ref{sec:refinements}.
\par
We keep the same format for the measurements of control inactive galaxies, with a short example shown in Table \ref{tab:A_galaxy}. In the complete version, we include the photo-z and stellar mass measurements from \cite{kawinwanichakij2021hyper}. The asymmetry measurement results are appended from the 9th column to the 73th column.



\bibliographystyle{mnras}
\bibliography{agn_morph} 




\appendix
\section{Refinement of asymmetry measurements} \label{sec:refinements}
\begin{figure*}
\begin{centering}
\includegraphics[width=0.95\textwidth]{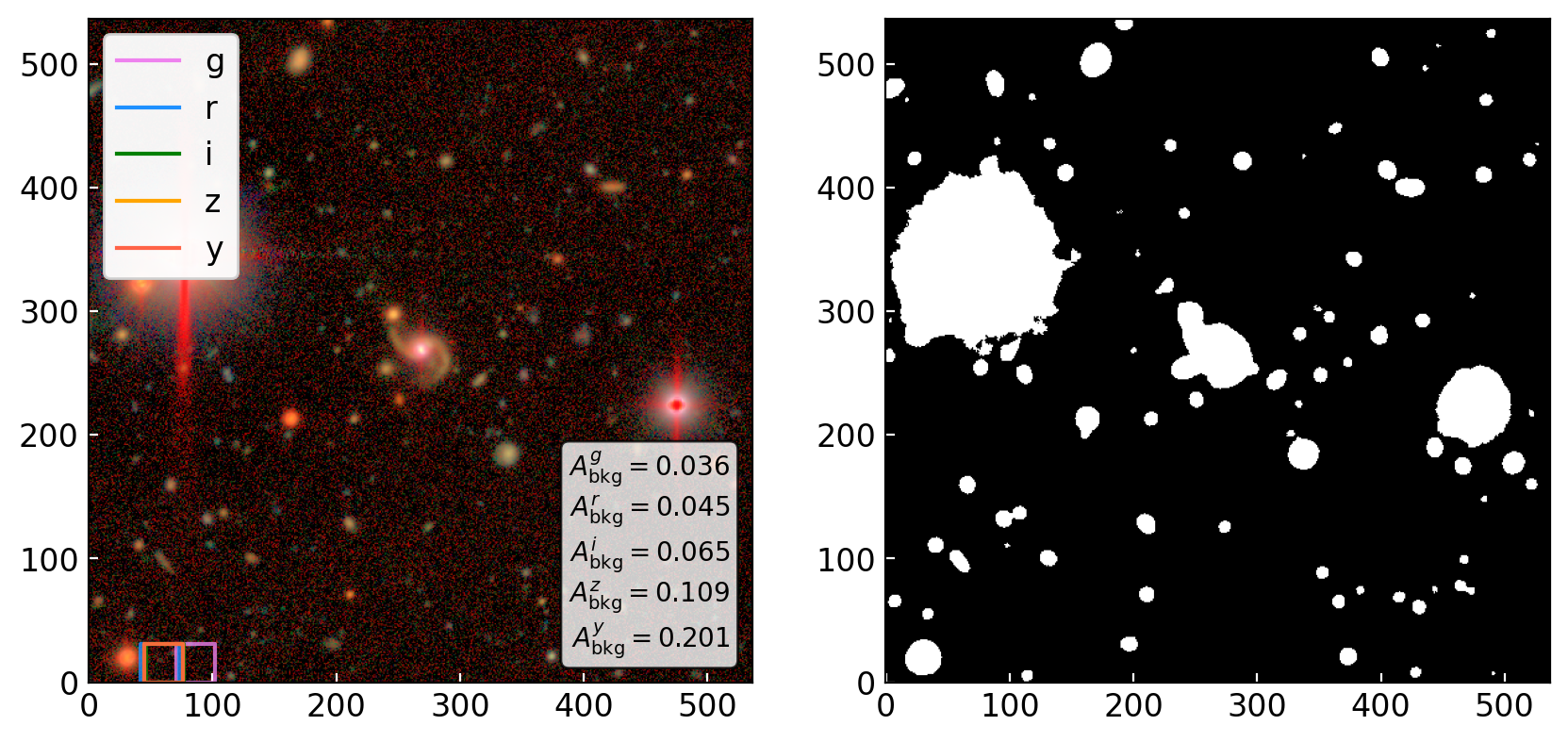}
\end{centering}
\caption{Proof of the "sky box" method to measure the sky asymmetry. Left panel: the five-band colored raw image of the quasar. The sky boxes are selected at empty spaces where there is no source detected. The sky asymmetry values per pixel are labeled. Right panel: $i$-band mask for the cutout image generated from a simple ``cool" mode detection on the raw data. \label{fig:bkg_proof}}
\end{figure*}

\begin{figure*}
\begin{centering}
\includegraphics[width=0.95\textwidth]{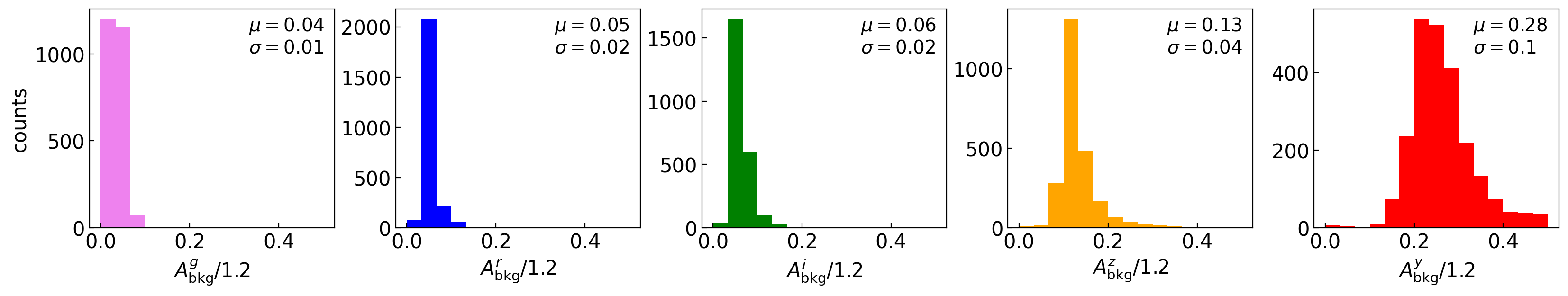}
\caption{Distribution of background asymmetry in all five bands ($A_{\rm bkg}^{\{g/r/i/z/y\}}/1.2$) for the quasar hosts that have reliable measurements. Mean ($\mu$) and standard deviation ($\sigma$) are shown for each band. We take $\sqrt{2}\sigma$ as the typical uncertainty levels of our asymmetry measurements on sources. \label{fig:sky_hist}}
\end{centering}
\end{figure*}

We first demonstrate the ``sky box" method in \textsc{statmorph} that we applied in Section \ref{subsec: CAS}. Figure \ref{fig:bkg_proof} (left panel) is a five-band colored image of quasar generated in the same way as Figure \ref{fig:QA}. With a raw image with size of $90 \arcsec \times 90 \arcsec$, the mask for sky asymmetry measurements (Figure \ref{fig:bkg_proof} right panel) is made from a simple ``cool" mode detection (Section \ref{subsec: MD}) without any deblending. Thus, all the sources that have at least 5 continuous pixels above $1\sigma$ threshold are masked. We make this mask for all five bands independently and the sky box is also drawn independently (colored squares in Figure \ref{fig:bkg_proof} left panel). They do not necessarily take the same area of the sky. We show the sky asymmetry values per pixel on the lower right side. 
\par
In Figure \ref{fig:sky_hist}, we show the distribution of background asymmetry measurements for all reliable samples. The standard deviation $\sigma$ can be considered as the typical uncertainty value for the background in each band. We can assume that the object also has the same level of uncertainty. Then according to function \ref{eq:A}, the uncertainty of the source measurement can be combined as $\sqrt{2}\sigma$. Generally, redder bands have higher sky asymmetry values and larger standard deviation than bluer bands, as the background noise increase from optical to NIR. 
\par
\begin{figure*}
\begin{centering}
\includegraphics[width=0.95\textwidth]{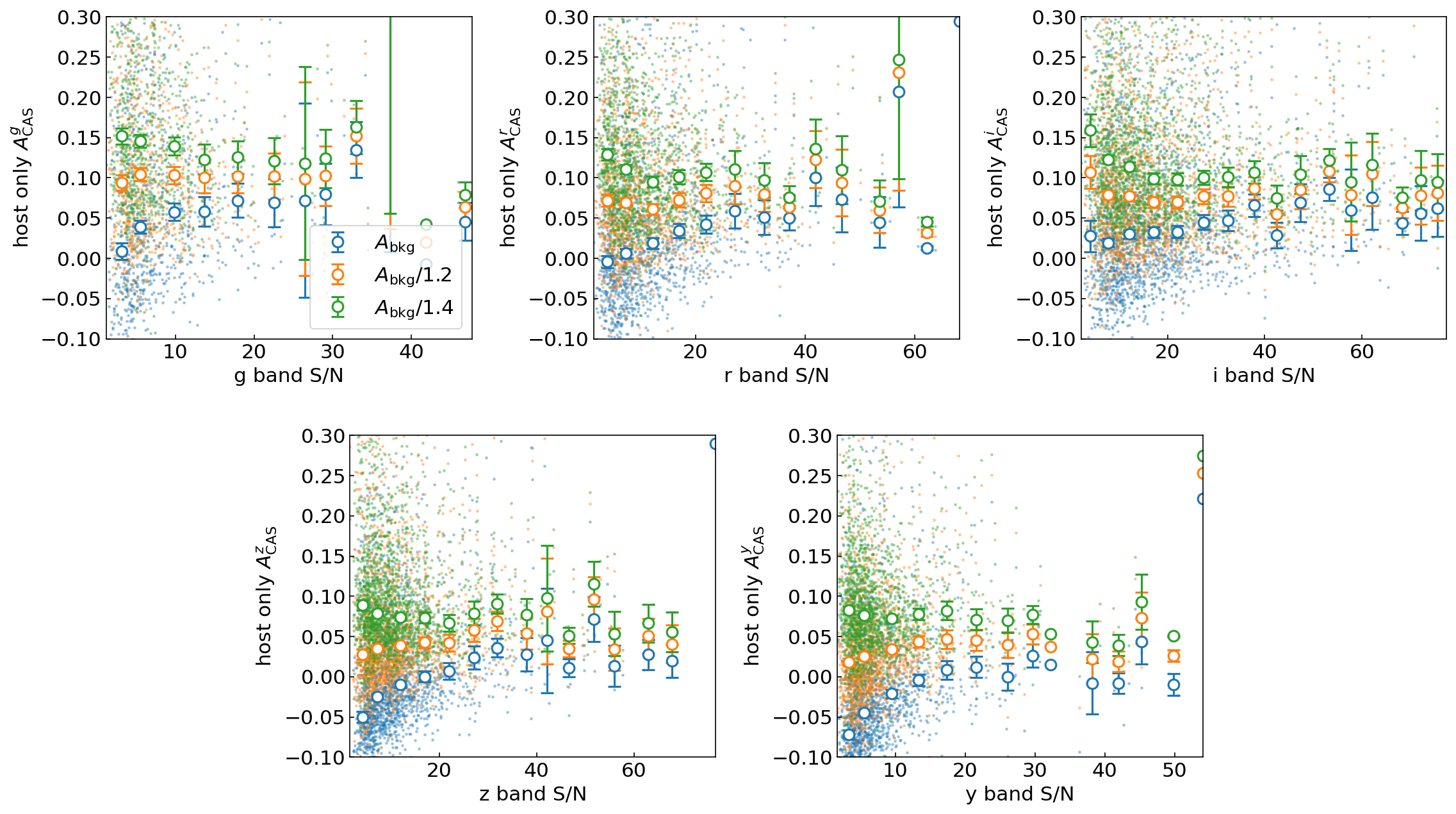}
\end{centering}
\caption{Comparison of $A_{\rm CAS}$ measured on ``host only" image via subtracting $A_{\rm bkg}$ (blue points), $A_{\rm bkg}/1.2$ (blue points), or $A_{\rm bkg}/1.4$ (blue points). The data are binned in S/N with a range of four for $g$ and $y$ bands, a range of five for $r$, $i$, and $z$ bands. Median values with mean $\sigma$ uncertainties are shown for each bin. \label{fig:noise_sub}}
\end{figure*}

\cite{thorp2021towards} pointed out that this method may over-subtract the sky asymmetry and lead to an underestimation of the real asymmetry value of the source. This is because the second item of Equation \ref{eq:A} takes the absolute value of $B_0 - B_{180}$, even a random noise field will have a positive value. This also explains why the noisier redder bands have larger sky asymmetry values. \cite{thorp2021towards} randomly selected 1000 galaxies from the IllustrisTNG simulation \citep{pillepich2018simulating} with $M_* \geq 10^{9} M_{\odot}$ at $z=0$. They added observational effects including the resolution and noise to the simulated galaxies and studied how well \textsc{statmorph} can reproduce the intrinsic asymmetry values. Their first main finding is that, instead of subtracting the whole $A_{\rm bkg}$ (they refer to it as $A_{\rm noise}$), dividing the value by 1.2 better recovers the intrinsic asymmetry. We tested with our $A_{\rm CAS}^i$ measurements on ``host only" images via subtracting the whole $A_{\rm bkg}$, or $A_{\rm bkg}/1.2$, or $A_{\rm bkg}/1.4$ (Figure \ref{fig:noise_sub}). We do not know the intrinsic asymmetry of our data, but since this modification aims to correct the bias caused by S/N, we would expect a successful correction to flatten the asymmetry values concerning S/N (similar to their Figure 4). In that sense, we roughly reproduced their curves at S/N$<$80, in which the subtraction of $A_{\rm bkg}/1.2$ has the stablest $A_{\rm CAS}$ over the whole S/N range for all five bands. While subtracting $A_{\rm bkg}/1.4$ and $A_{\rm bkg}$ tend to overestimate and underestimate $A_{\rm CAS}$ at S/N$<$15. Also, subtracting $A_{\rm bkg}/1.2$ instead of $A_{\rm bkg}$ significantly reduced the fraction of negative $A_{\rm CAS}$ from $19\% \sim 68\%$ to $3\% \sim 23\%$ in these bands.\par

\begin{figure*}
\begin{centering}
\includegraphics[width=0.65\textwidth]{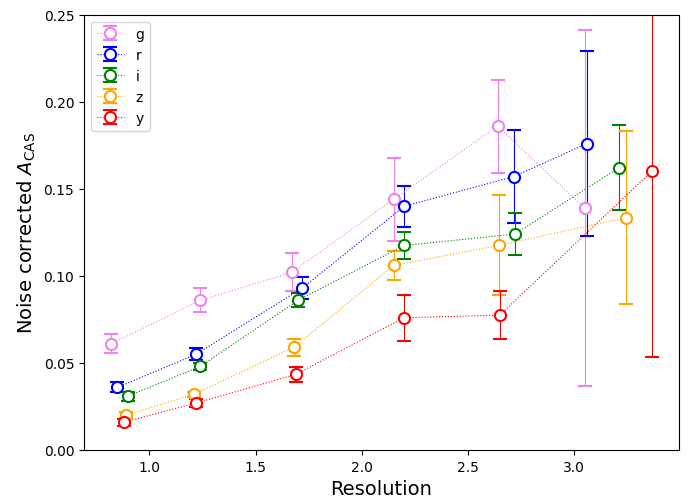}
\caption{The noise-corrected $A_{\rm CAS}$ of quasar hosts binned in resolution ($R_{\text {1/2}} / \mathrm{FWHM}$). Median values are shown with mean $\sigma$ errors in all five bands. A clear trend exists that $A_{\rm CAS}$ increases with resolution, the reason of which could either be technical or physical. We do not apply the resolution correction to our results given its strong degeneracy with S/N. Nevertheless, the offset between the bands at almost all resolution bins in this plot indicates that the band dependence of $A_{\rm CAS}$ should attribute to some other reasons, at least not only the resolution. \label{fig:Resolution_A}}
\end{centering}
\end{figure*}

Besides S/N, the low resolution may also smooth out an image and lead to underestimated asymmetry values. To address this, we calculated the resolution of each quasar host galaxy in each band as $\mathrm{resolution}=R_{\text {1/2}} / \mathrm{FWHM}$ where $R_{\text {1/2}}$ is the circular half-light radius calculated by statmorph.source\_morphology.rhalf\_circ, and FWHM is calculated using galight.tools.measure\_tools.measure\_FWHM on the input HSC PSF images. Figure \ref{fig:Resolution_A} shows the median $A_{\rm CAS}$ in resolution bins for all five bands. There is a clear trend that the values decrease towards low resolution. However, it is unclear whether this trend is purely caused by resolution since it can be entangled with S/N. Also, it could be physical effect since disturbed galaxies are more likely to have larger $R_{\text {1/2}}$. \cite{thorp2021towards} also performed a fitting function to correct for the resolution issue, but the scatter increases significantly compared to the results with only noise corrections. Therefore, we decided to adopt the $A_{\rm bkg}/1.2$ noise correction and not the resolution correction to our results. In the mean time, we claim from this figure that the band differences are independent of resolution thus supportive of the results given in Section \ref{subsec:band}.

\section{Control Galaxy measurements} \label{sec:control_G}
For the inactive galaxies (control sample), we proceed with the image analysis as done for the quasar images as described in Section \ref{sec:methods}. Here, the difference is that there is no point source to subtract.
All components within the image are modeled with \sersic\ profiles at the same depth (200 iterations) for the PSO algorithm. Then these components, with the exception of the central galaxy, are removed to produce the ``host only" frame. While the control galaxies are not "hosting" a quasar, we keep the name as used for the quasar frames. As well, the processes of making segmentation masks and measuring asymmetry are the same as done for the quasars. Figure \ref{fig:QA_G} shows an example of the \textsc{statmorph} measurements on one of the galaxies.\\

\begin{figure*}
\begin{centering}
\includegraphics[width=0.95\textwidth]{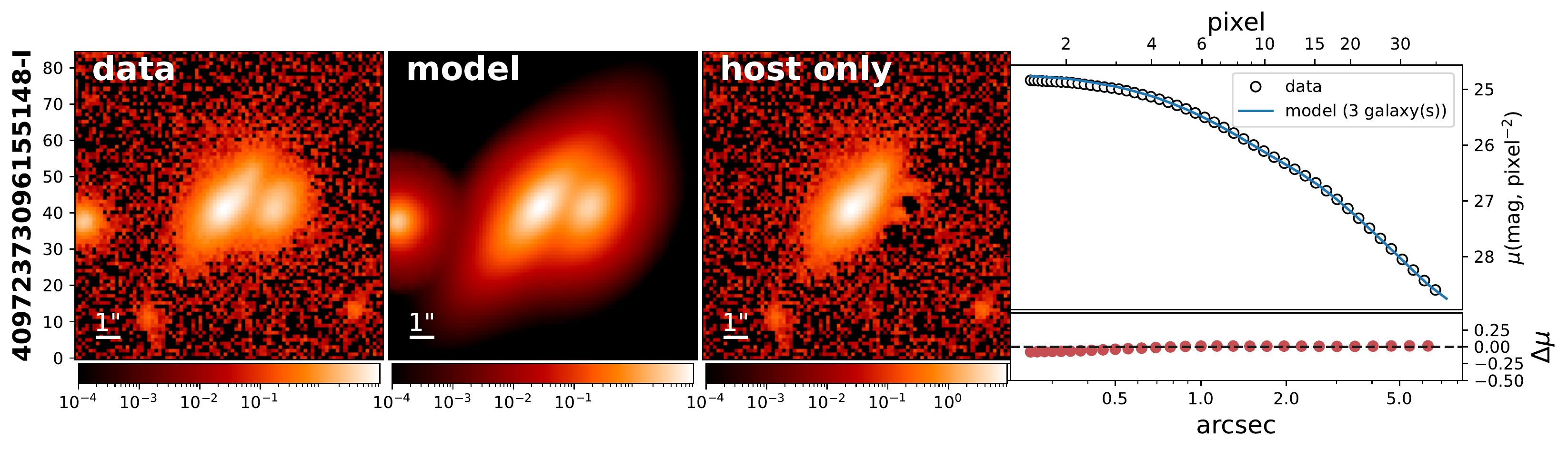}
\end{centering}
\caption{Similar to Figure \ref{fig:fit_comp}, but for control galaxy, HSC ID 40972373096155148, RA=133.87344, Dec=-0.08921. In case of control galaxies, data (first panel) play the role of "host+comp" frame. \label{fig:fit_comp_G}}
\end{figure*}

\begin{figure*}
\begin{centering}
\includegraphics[width=0.95\textwidth]{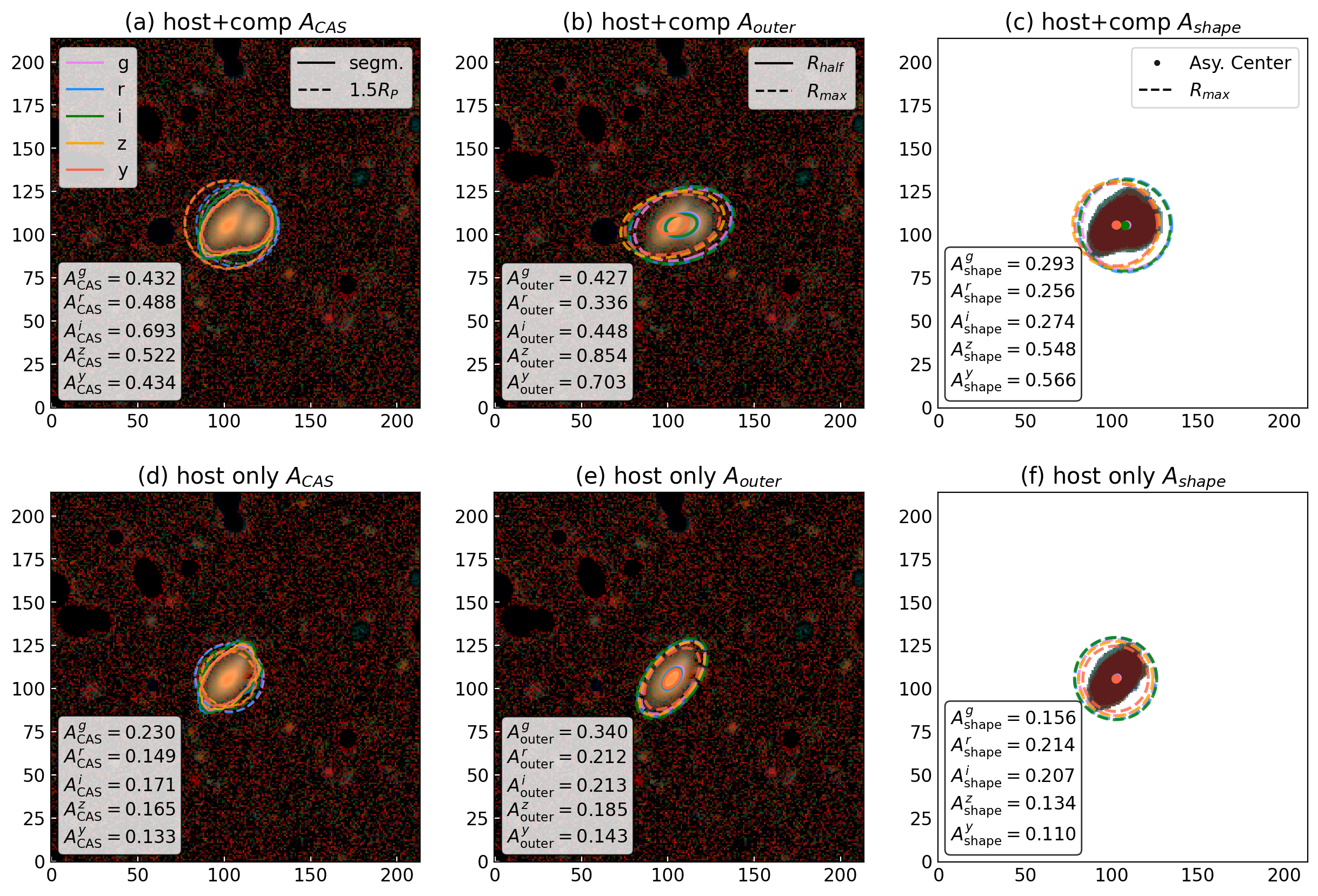}
\end{centering}
\caption{Similar to Figure \ref{fig:QA}, but for the control galaxy in Figure \ref{fig:fit_comp_G}.\label{fig:QA_G}}
\end{figure*}

\begin{figure*}
\begin{centering}
\includegraphics[width=0.65\textwidth]{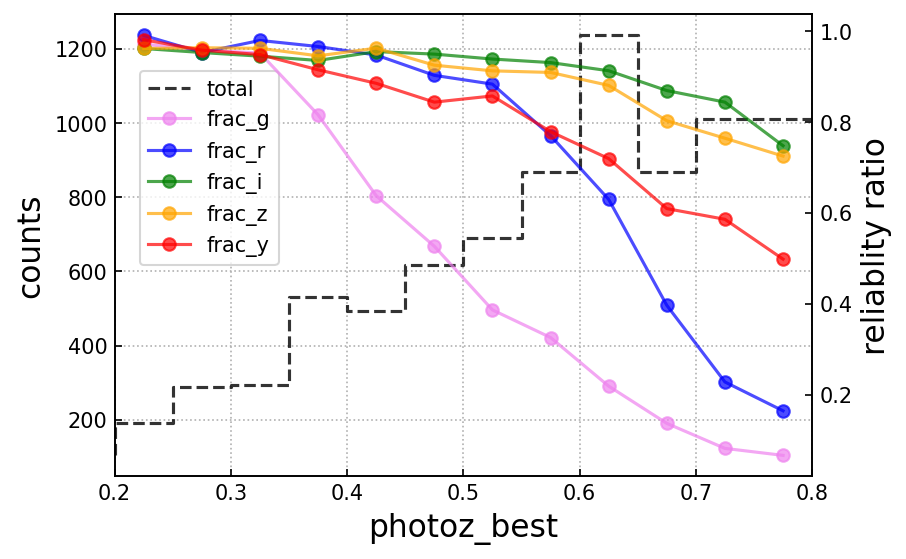}
\caption{Similar to Figure \ref{fig:completeness_Q}, but for control galaxies.
\label{fig:completeness_G}}
\end{centering}
\end{figure*}

We then select reliable measurements from the control galaxies in the same manner as for the quasars. The corresponding numbers of reliable objects in every band are \textcolor{black}{g, 2363; r, 4288; i, 6428; z, 6200; y, 5199}. Figure \ref{fig:completeness_G} shows the counts in each redshift bin and the corresponding reliability ratios in each band, which generally shows the same trends as the quasar hosts.

\section{Robustness of the correlations} \label{subsec:uncertainties}
\begin{figure*}
\begin{centering}
\includegraphics[width=0.95\textwidth]{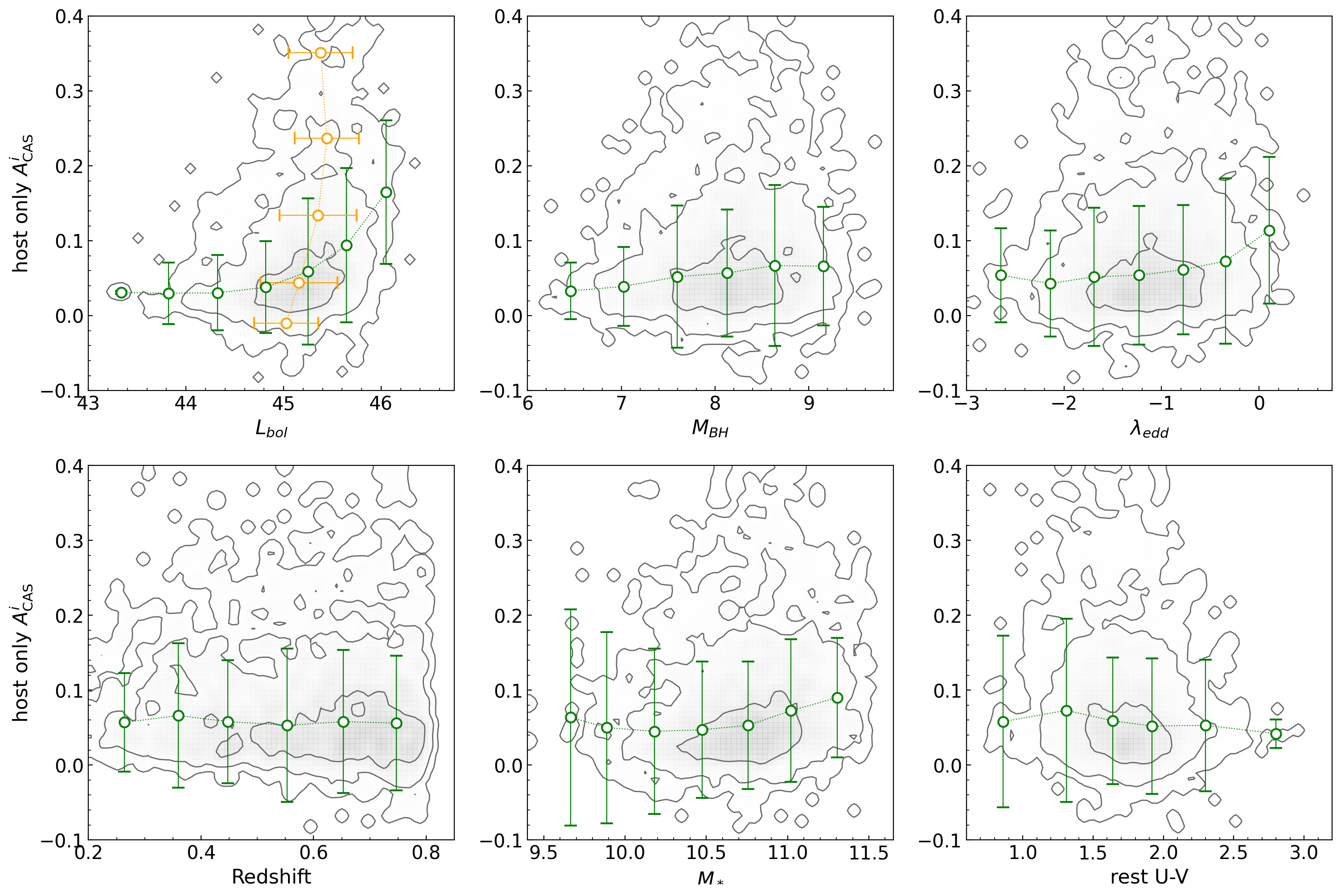}
\caption{Similar to Figure \ref{fig:host_A_evolution_single}, but include the sources rejected by \textsc{statmorph}, i.e., flag=1. We found the results (trends) are consistent in both cases.
\label{fig:host_A_evolution_single_all}}
\end{centering}
\end{figure*}

\textcolor{black}{We test the robustness of the reported correlations from two aspects. In Section \ref{subsec:qualification}, we rejected sources that are flagged by \textsc{statmorph}. It is difficult to figure out the exact reason of each rejection since only one flag value is provided. We visually checked some of the rejected sources and did not find significant issues in most of the cases. Therefore, we put those sources back and reproduced Figure \ref{fig:host_A_evolution_single} as Figure \ref{fig:host_A_evolution_single_all}. This increases the reliable source number from 1987 to 2409. Now the rejection is only made by our deblending and mask pipeline. For the rejected 15 sources, 10 are rejected due to bad subtraction residuals, which lead to a failure in source detection; 3 are rejected due to contamination from large foreground sources; 1 is rejected due to artifact; 1 is rejected due to no-detection of the host. None of these rejected sources has merger features, and the median bolometric luminosity is 45.4, similar to the median of entire sample within $1\sigma$ ($45.2\pm0.4$). Therefore, our rejections do not bias to specific type of objects. In Figure \ref{fig:host_A_evolution_single_all}, we find consistent trend as in Figure \ref{fig:host_A_evolution_single}, thus as a proof that our results are robust with different rejection criteria.} 
\par
\begin{figure*}
\begin{centering}
\includegraphics[width=0.95\textwidth]{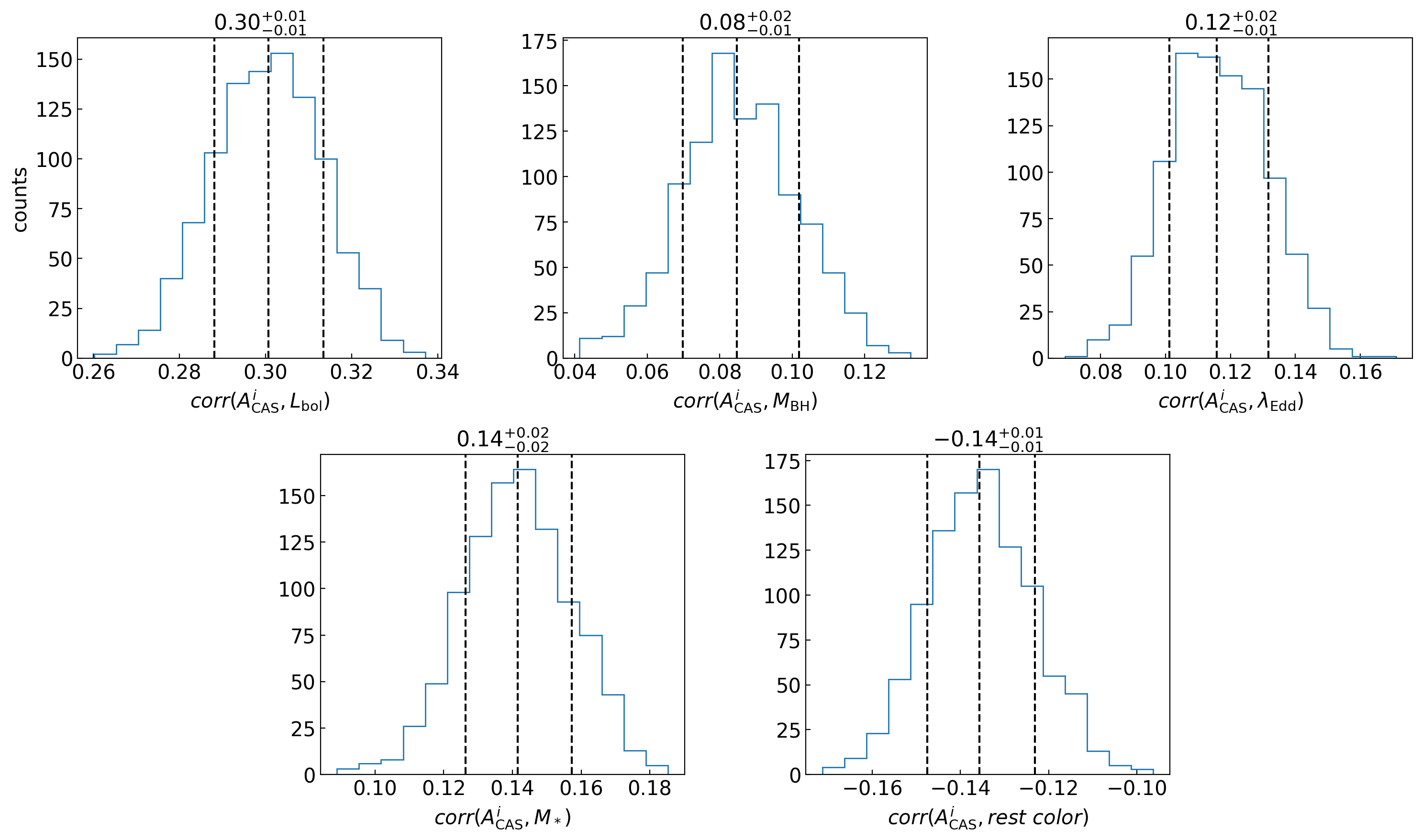}
\end{centering}
\caption{Distribution of the spearman correlation coefficients between $A_{\rm CAS}^i$ and other parameters over 1000 time Monte Carlo simulations for all the quasars that have reliable measurements of $A_{\rm CAS}^i$. The 16th, 50th, and 84th percentiles are labeled by the dashed vertical lines and the numbers on top of each panel. \label{fig:corr_distribution}}
\end{figure*}
We then tested how the uncertainty of the parameters affect the correlations reported in Figure \ref{fig:corr_matrix} using Monte Carlo simulations. We assumed gaussian distribution of the six parameters: $A_{\rm CAS}^i$, $L_{\rm bol}$, $M_{\rm BH}$, $\lambda_{\rm edd}$, $M_*$ and rest-frame U-V color. The distribution in Figure \ref{fig:sky_hist} can be used to estimate the uncertainty level of $A_{\rm CAS}$ as $\sqrt{2}\sigma$ combining the object term and background term. For $i$-band, we take it as $\sqrt{2}\times0.02=0.03$. The uncertainty on $L_{\rm bol}$ is mainly attributed to the bolometric correction (BC) factor. \cite{richards2006spectral} reported a $\sigma$ of $\sim$ 2 for $\mathrm{BC}_{5100 \text{\AA}}$, which converts to $\sim$ 0.1 in log $L_{\rm bol}$ according to Equation \ref{eq:Lbol}. The uncertainty level of $L_{5100}$ is $\sim$ 0.01 according to the \cite{rakshit2020spectral} catalog, thus we ignore it and only considered the uncertainty on BC. The error of $M_{\rm BH}$ is provided in the \cite{rakshit2020spectral} catalog, which is mainly caused by the spectral fitting and inherent systematics of the viral method. We use these $M_{\rm BH}$ errors as the $\sigma$ for each of our quasar. We then estimated the uncertainty level of $\lambda_{\rm edd}$ via combining the $\sigma$ of $L_{\rm bol}$ and $M_{\rm BH}$ in Equation \ref{eq:R_edd}. The errors ($\sigma$) on $M_*$ is provided by \cite{li2021sizes}, which is a result of SED fitting using CIGALE. \cite{li2021sizes} also compared their rest frame U-V color measurements to the values of CANDELS galaxies, and found a median difference of 0.03. We used this as the $\sigma$ on the U-V color. \par
We perform this simulation for all quasars that have a valid $A_{\rm CAS}^i$ measurements. For each iteration, we calculated the spearman correlation coefficients between $A_{\rm CAS}^i$ and the other parameters. The distribution of the coefficients in 1000 iterations is shown in Figure \ref{fig:corr_distribution} with their 16th, 50th, and 84th percentile values labelled as dashed lines. We find that adding these uncertainties reduces the correlation coefficient that we have reported in Figure \ref{fig:corr_matrix} by $\sim 20\%$, which does not affect our main results.



\bsp	
\label{lastpage}
\end{document}